\documentclass[twocolumn]{aastex631}

\usepackage{amsmath,ulem}
\usepackage{enumitem}

\usepackage{color, hyperref}
\hypersetup{colorlinks = true, linkcolor = blue, citecolor = blue, filecolor = linkcolor, urlcolor = blue}

\usepackage{float}
\usepackage{graphicx}

\newcommand{\Tf}{T_\mathrm{eff}}
\newcommand{\Dn}{\Delta\nu}
\newcommand{\eps}{\epsilon}

\begin{document}

\title{\Large \bf Asteroseismology applied to constrain structure parameters of $\delta$ Scuti stars}

\author{Subrata Kumar Panda}
\affiliation{Tata Institute of Fundamental Research, Colaba, Mumbai 400005, Maharashtra, India}

\author{Siddharth Dhanpal}
\affiliation{Tata Institute of Fundamental Research, Colaba, Mumbai 400005, Maharashtra, India}

\author{Simon J. Murphy}
\affiliation{Centre for Astrophysics, University of Southern Queensland, Toowoomba, QLD 4350, Australia}

\author{Shravan Hanasoge}
\affiliation{Tata Institute of Fundamental Research, Colaba, Mumbai 400005, Maharashtra, India}

\author{Timothy R. Bedding}
\affiliation{Sydney Institute for Astronomy (SIfA), School of Physics, University of Sydney, NSW 2006, Australia}

\begin{abstract}
Asteroseismology is a powerful tool to probe the structure of stars. Space-borne instruments like CoRoT, Kepler and TESS have observed the oscillations of numerous stars, among which $\delta$ Scutis are particularly interesting owing to their fast rotation rates and complex pulsation mechanisms. In this work, we inferred \textit{model-dependent} masses, metallicities and ages of 60 $\delta$ Scuti stars from their photometric, spectroscopic and asteroseismic observations using least-squares minimization. These statistics have the potential to explain why only a tiny fraction of $\delta$ Sct stars pulsate in a very clean manner. We find most of these stars with masses around $1.6 ~{\rm M_\odot}$ and metallicities below $Z=0.010$. We observed a bimodality in age for these stars, with more than half the sample younger than 30 Myr, while the remaining ones were inferred to be older, i.e., hundreds of Myrs. This work emphasizes the importance of the large-frequency separation ($\Dn$) in studies of $\delta$ Scuti stars. We also designed three machine learning (ML) models that hold the potential for inferring these parameters at lower computational cost and much more rapidly. These models further revealed that constraining dipole modes can help in significantly improving age estimation and that radial modes succinctly encode information pertaining to stellar luminosity and temperature. Using the ML models, we also gained qualitative insight into the importance of stellar observables in estimating mass, metallicity, and age. The effective surface temperature $\Tf$ strongly affects the inference of all structure parameters and the asteroseismic offset parameter $\eps$ plays an essential role in the inference of age.
\end{abstract}

\section{Introduction} \label{intro}
Asteroseismic observations from TESS (\citealt{tessmission}) and Kepler (\citealt{keplermission}) have shed light on the dynamics and interiors of thousands of pulsating stars (\citealt{Paparo2019, Bowman2020, Aerts2021, Kurtz2022}). A sizeable fraction of the pulsating class from these missions are $\delta$ Scuti stars (\citealt{Bowman2018, Guzik2021}), which lie at the junction of the instability strip and the main sequence (MS) band in the Hertzsprung-Russell (HR) diagram (\citealt{spectraltype}), providing direct views of both these classes. These are low-to-intermediate mass ($1.5-2.5 M_\odot$) main-sequence variables with spectral types ranging from A-F (\citealt{Bowman2018, Murphy2021, Kurtz2022}). High amplitude delta Scuti stars (HADS) are used as standard candles (\citealt{McNamara_2007}), and in assessing the metallicities and ages of stellar clusters (\citealt{Murphy2022}), in turn enabling Galactic archaeology.

Asteroseismology is a powerful tool which can be used to estimate various structure parameters such as mass, composition, and age on different pulsating classes, e.g., stochastic oscillators (\citealt{Chaplin2013, Hekker2017, RGBpaperV, RGBpaperM0, RGBpaperM, rgb-dnn}), $\gamma$-Doradus stars (\citealt{gmd611paper, gmd-dnn}) and high-mass coherent oscillators (\citealt{highM-dnn}), among others. Some efforts have been applied to similar studies of $\delta$ Sct stars (\citealt{lim_prog1, lim_prog4, lim_prog3, lim_prog2}). In this paper, we develop a methodology to measure structure parameters - mass, metallicity, and ages of $\delta$ Sct stars. These measurements can in principle constrain the metallicities and ages of the host open clusters from which these stars formed and were dispersed before escaping to become field stars.

$\delta$ Sct stars predominantly exhibit low radial-order pulsations, with high-frequency pressure and low-frequency gravity  modes, roughly separated around $5 ~\mathrm{d}^{-1}$, although this depends on $\Tf$ to leading order (\citealt{Moya}). The pressure modes are mainly driven by the $\kappa$ mechanism (\citealt{kappamech}) in the Helium ionization zone. They primarily propagate in the stellar envelope and probe the near-surface regions.

Although we have precise estimates of the luminosity ($L$), effective temperature ($\Tf$) and pulsation frequencies, the seismology and the parameter inference of $\delta$ Sct stars is  challenging because of the following reasons: (a) they have low radial order \textit{p}-modes and \textit{g}-modes, where asymptotic theory fails, (b) many stars exhibit fast rotation, leading to ellipsoidal deformation (\citealt{Reese2006}), (c) complex mode-selection mechanisms (\citealt{mode-selection}) influencing the observed spectra, and (d) appearance of island modes (\citealt{Reese2006}), chaotic modes (\citealt{chaotic_real_star}) etc. making it very difficult to characterize $\delta$ Sct spectra.

Finding regular patterns in $\delta$ Sct spectra is very helpful for carrying out asteroseismology. Although some $\delta$ Sct stars had been previously reported to  show regular frequency spacings (\citealt{Matthews, reg2009, regular_ds_MOST, corot_ds_regularity, measure_density_ds}), a larger ensemble of such stars was found by \citet{nature}, who identified 60 $\delta$ Sct stars from {\it TESS} and {\it Kepler} exhibiting regular pulsation patterns. They were able to identify some modes, label their radial orders and deduce the large-frequency separation ($\Dn$) and {\it p}-mode offset ($\eps$). 
% added by TimB
Many more examples of high-frequency patterns in $\delta$ Sct stars have subsequently been identified \citep[e.g.,][]{Murphy2020-lambda-boo, Murphy2021, Hasanzadeh2021, LeDizes2021-Altair, KahramanAlicavucs2022-TZ-Dra, Murphy2022}.
Here, we use seismic parameters to infer stellar structure through the application of neural networks and other techniques. As the relationships between observables such as luminosity, temperature, pulsation frequencies and structure parameters can be highly non-linear, neural networks can potentially be useful in building a model connecting them all. In addition, a well-trained neural network infers these structure parameters substantially faster than conventional methods such as MCMC, making it a capable method for ensemble studies.

At the core of this work lies the simulated models of $\delta$ Sct stars, which we briefly describe in section \ref{evol-grid}. Following that we elaborate on developing three different machine learning methods in Section \ref{method-ml}, which in principle could have inferred mass, metalicity, and ages of $\delta$ Sct stars. However, none of the above methods worked uniformly well over all the stars in our sample because of various limitations. Yet, the networks helped us in gaining qualitative insights into importance of different observables, which we present alongwith each method.
Since our goal was to characterize the 60 $\delta$ Scutis from \citealt{nature}, we finally deployed a grid-search based least-squares minimization technique in Section \ref{method4}, with which we were able to infer ($M, Z, \tau$) parameters for all stars. We end the article with presenting the statistics of these inferences. While we have presented all the machine-learning based methods for the sake of completeness, we found that the least-squares fitting method performed best and therefore regard its output as our final result.

\section{Grid of stellar models} \label{evol-grid}
We have used the model grid presented in \cite{Murphy2023} which contains more than 800,000 stellar models. However since we aimed for training the networks on $\delta$ Sct stars, we extracted 524,247 models based on a simpler criterion of 6500 K $\le \Tf \le$ 10000 K. MESA inlists used for the grid computation and the corresponding output models are available as supplementary files in \citealt{Murphy2023}. This grid covers a wider span of mass ($M$) and metallicity ($Z$) over the ranges $[1.3 - 2.2] ~ M_\odot$ and $[0.002 - 0.026]$, with $0.1 M_\odot$ ($M$) and 0.002 ($Z$) resolution respectively.

These models were evolved using the stellar evolution code \textsc{mesa} (\citealt{MESA2011, MESA2013, MESA2015, MESA2018, MESA2019}) and corresponding pulsation frequencies were calculated using \textsc{gyre} (\citealt{GYRE2013, GYRE2017, GYRE2020}) -- which in turn provides the large-frequency separation ($\Dn$) and offset parameter ($\eps$).

The computed eigenmodes comprise p, g and mixed modes oscillating at frequencies between 0 to 95 $d^{-1}$ and at harmonic degrees of $\ell$ = 0 and 1. Since the majority of $\delta$ Sct stars pulsate in radial and dipole modes, and higher-degree modes are not observable due to geometric cancellation, the availability of modes with degree $\ell$ = 2 or higher is not necessary for our purpose. All the computed modes correspond to non-rotating $m=0$ components. $\Dn$ and $\eps$ were calculated by fitting the asymptotic relation to radial ($\ell=0$) p-modes with order ($n_{\rm pg}$) larger than 5.

We used this grid to train the neural networks to infer structure parameters from observables and asteroseismic quantities.

\begin{figure}[tb]
    \centering
    \includegraphics[width=0.74\columnwidth]{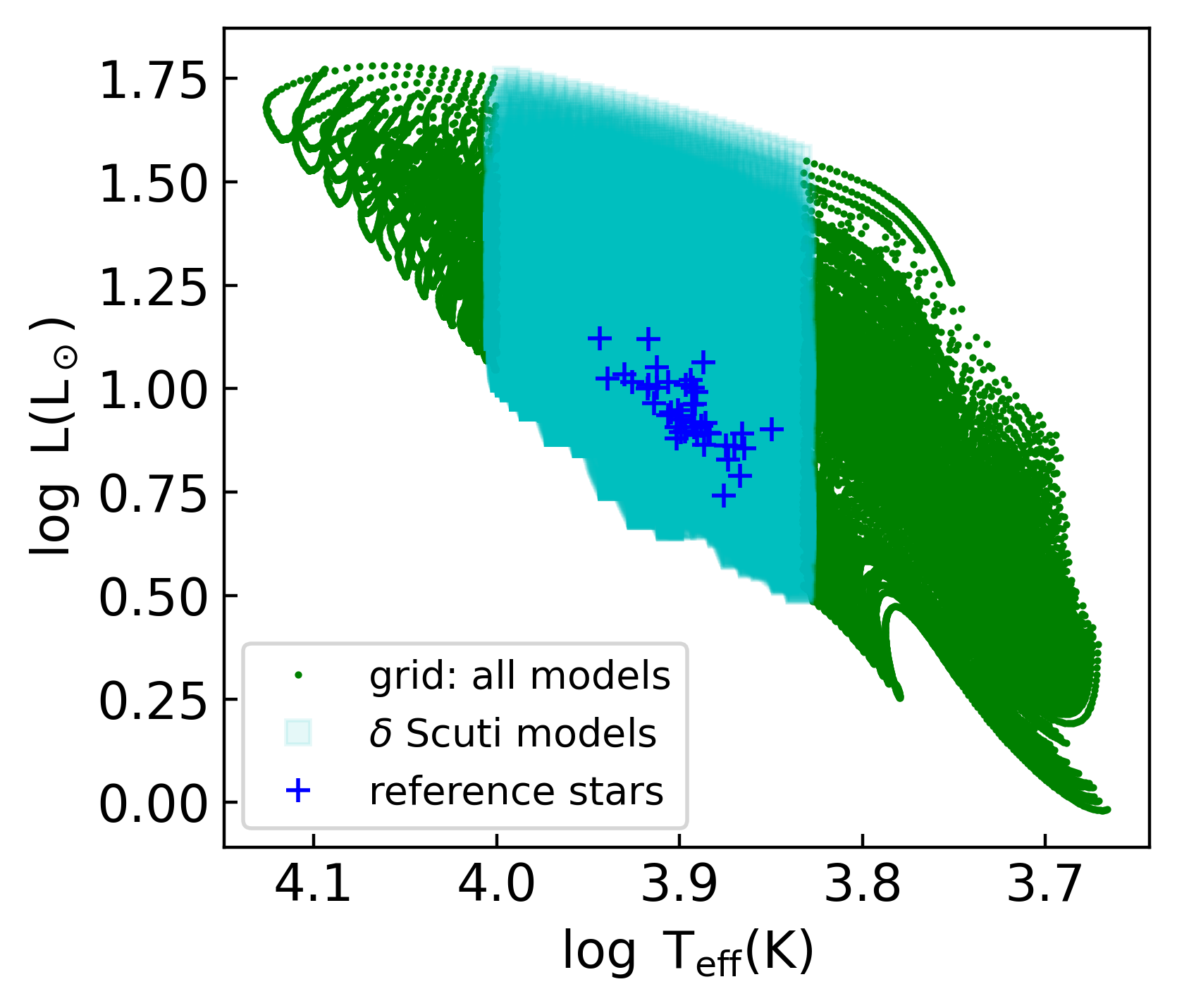}
    %\captionsetup{labelfont={color=blue, bf}}
    \caption{HR diagram showing the positions of all synthetic stellar models, with $\delta$ Sct stars present in the solid patch. The $\delta$ Sct targets of this paper are denoted using `+' symbols.}
    \label{grid-in-HR}
\end{figure}

\begin{figure}[tb]
    \centering
    \includegraphics[width=0.7549\columnwidth]{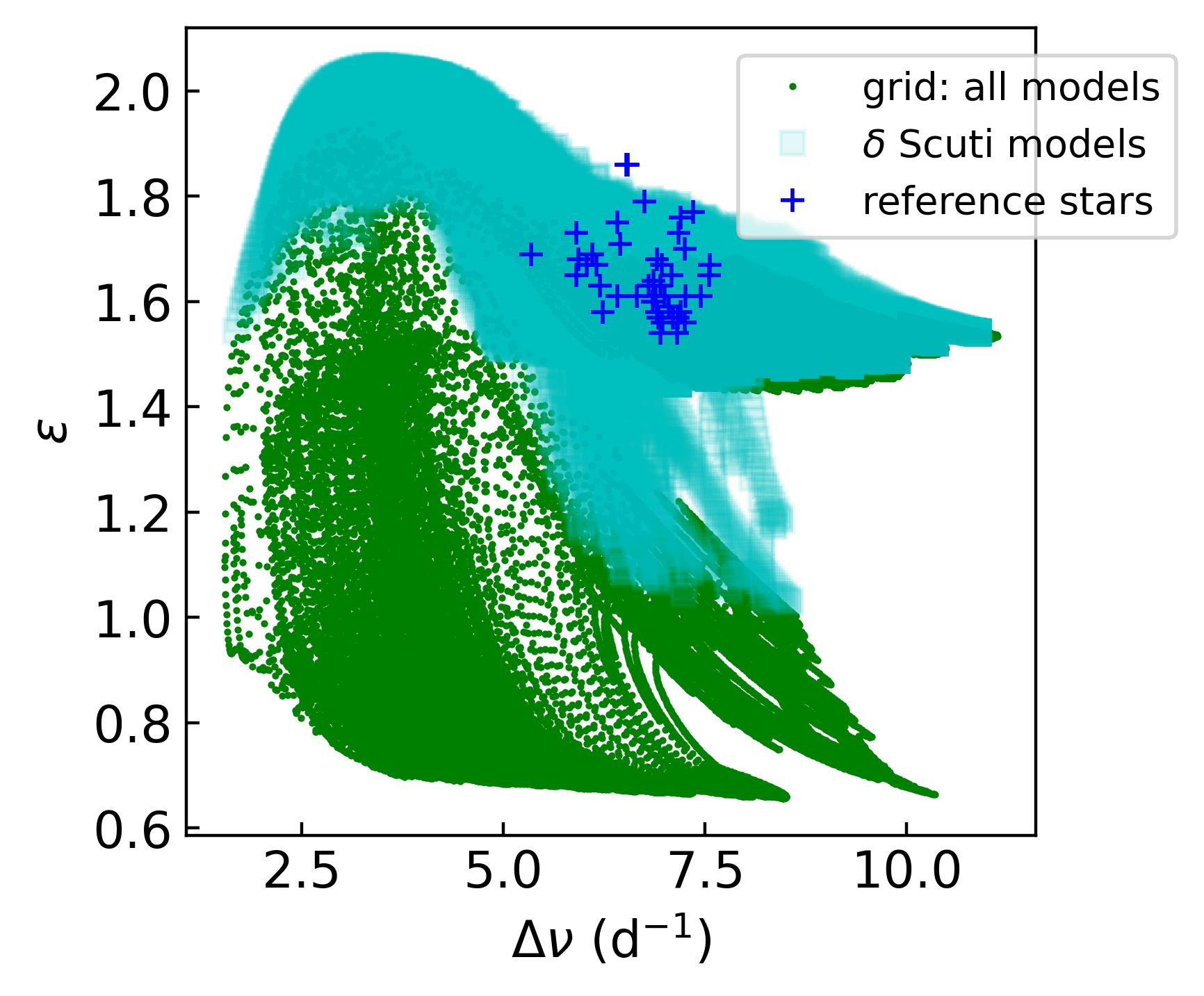}
    %\captionsetup{labelfont={color=blue, bf}}
    \caption{$\Dn - \eps$ diagram denoting the positions of all synthetic stellar models, with $\delta$ Sct stars present in the solid patch. The $\delta$ Sct targets of this paper are denoted using `+' symbols.}
    \label{grid-in-Dnueps}
\end{figure}

The HR diagram (Figure \ref{grid-in-HR}) shows the stellar models, including the potential $\delta$\,Sct models ($3.83 \le \log\Tf \le 4$) on this HR diagram. For reference, we mark the target stars of this work using `+' symbols, which are taken from the observed $\delta$\,Sct stars in \citet{nature}. Figure~\ref{grid-in-Dnueps} contains the same models as Figure \ref{grid-in-HR}, but showing the asteroseismic $\Dn - \eps$ diagram. Both figures demonstrate that the phase space of our stellar grid is broad enough to represent the wide range of observed $\delta$\,Sct stars.

\section{Machine learning based methods} \label{method-ml}
The application of machine learning (ML) algorithms to the fitting of stellar models has shown great promise (\citealt{stoch-dnn, highM-dnn, rgb-dnn, scutt2023asteroseismology}).
For the work presented in this article, we have developed three different versions of neural networks which perform regression to infer the values of $M$, $Z$ and age ($\tau$). Our networks were developed using the Python libraries \textsc{tensorflow} (\citealt{tensorflow}) and \textsc{keras} (\citealt{keras}).

In each set of ML experiments, we randomly split the $\delta$ Sct model grid into `training' and `test' (validation) data at a $95:5$ ratio. The network used the training data to learn different features and compute the output. The training aimed to minimize the loss function by reducing the mean-squared error between the actual and predicted outputs (equation \ref{eq:mse}) through stochastic gradient descent.
\begin{equation} \label{eq:mse}
    \mathrm{MSE} = \sum_{i=1}^N \dfrac{(y^i_\mathrm{predicted}-y^i_\mathrm{true})^2}{N}
\end{equation}
%where $N$ is the number of training samples.

The training takes place through a back propagation technique which gradually optimizes the network parameters using the \textsc{Adam} optimizer (\citealt{adampaper}). To avoid the situation where networks memorize the dataset instead learning the inherent features, we equipped the output layer of each network with an L2 regularizer having regularizing parameter $\lambda = 10^{-6}$.

Once the training is over, we validated network accuracy on the unseen test data and also compared the network prediction against ground truth. If the accuracy and the comparison were deemed acceptable on the validation data, we concluded that the network was successfully trained without overfitting.

We used the cross-validation method (\citealt{cross-validation}) to yield a distribuion of 40 instances of inferences from trained network of which median, 16$^\mathrm{th}$ and 84$^\mathrm{th}$ percentiles were calculated to obtain most probable result, lower and higher uncertainties.

However, these three networks were not able to yield meaningful inferences over real stars due to several limitations discussed below. Nonetheless, they have the potential to provide qualitative insights on relative importance of different observables in the way (\citealt{bellinger}) assessed for main sequence stars. We performed this analysis for all the three networks. Physical interpretation of such analysis are also presented.

\subsection{Method 1: ML using seismic indices} \label{method1}
In this method we deployed three regression networks to infer $M$, $Z$ and $\tau$ from four input parameters: \{$L, \Tf, \Dn, \eps$\}. Although radius ($R$) is dependent on $L$ and $\Tf$, supplementing  it as an additional input helped the networks train quickly and achieve higher accuracy.

For the network to infer age ($\tau$), it typically requires prior knowledge of $M$ and $Z$: this is because each physical parameter ($\theta$) is a functions of $M$, $Z$ and $\tau$, which is why age inversion requires ($M$, $Z$) to be supplied as inputs.

$$\theta = f(M, Z, \tau) \implies \tau = \Tilde{f}^{-1} (M, Z, \theta)$$

\begin{figure}[tb]
    \centering
    \includegraphics[scale=0.24]{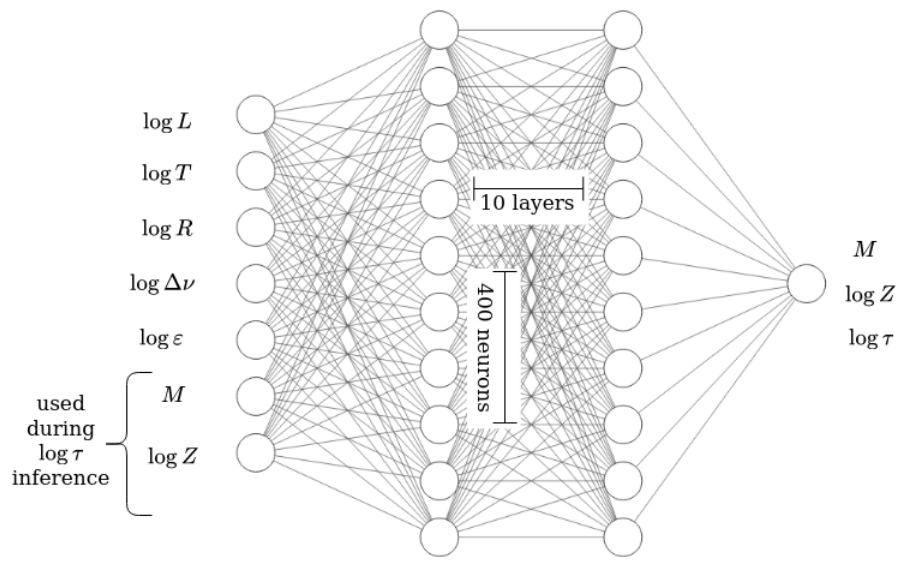}
    %\captionsetup{labelfont={color=blue, bf}}
    \caption{Architecture of the neural-network models.}
    \label{fig:model-architecture}
\end{figure}

The network architecture comprised of 1 input layer (with 5 to 7 neurons), 10 intermediate layers (of 400 neurons each), and 1 single-neuron output layer (Figure~\ref{fig:model-architecture}).
With the exception of the output layer, which has a {\it tanh} activation function, all other neurons were activated using the rectified linear unit (reLU) function.
Since we developed a few alternative methods, we label this procedure as \textit{Method-1} for easy reference.

\begin{figure*}[tb]
    \includegraphics[width=\textwidth]{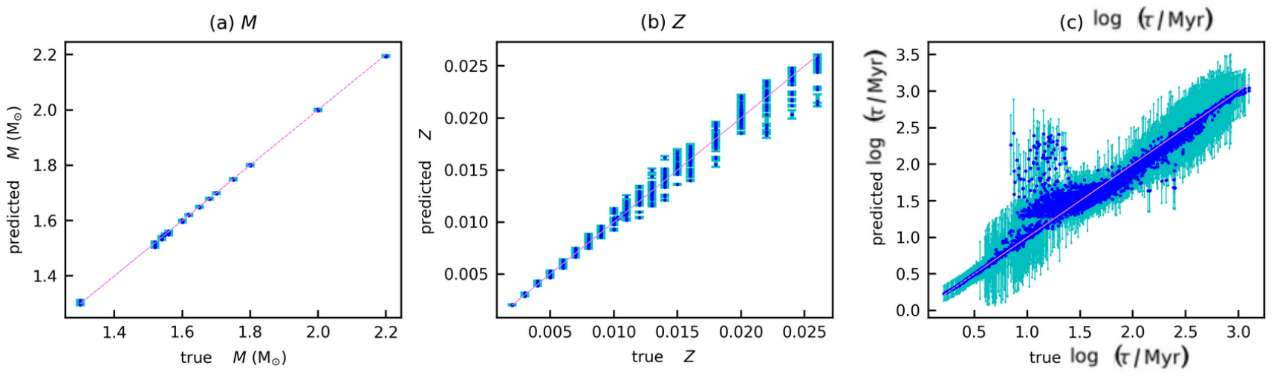}
    %\captionsetup{labelfont={color=blue, bf}}
    \caption{Plots showing network predictions for validation samples against their corresponding true values. Shown for (a) $M$, (b) $Z$, and (c) $\log\tau$. Uncertainties associated with predicted values are also shown.}
    \label{fig:validation}
\end{figure*}

\begin{figure*}[tb]
    \includegraphics[width=\textwidth]{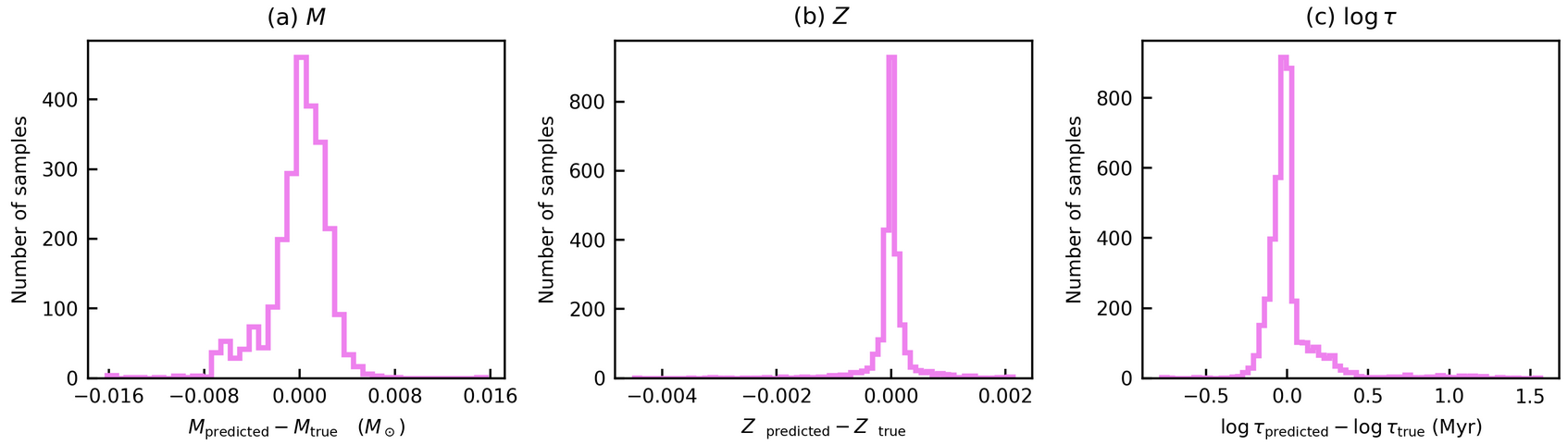}
    %\captionsetup{labelfont={color=blue, bf}}
    \caption{Distribution of errors between network predictions for validation samples and their corresponding true values: shown for (a) $M$, (b) $Z$, and (c) $\log\tau$.}
    \label{fig:validation_error}
\end{figure*}

We trained the networks and compared the network-predicted values for the validation data with the corresponding true values (Figure \ref{fig:validation}). The network was able to achieve high accuracy in predictions of $M$, and $Z$. Figure~\ref{fig:validation_error} shows the distributions of errors in our predictions on the validation data.

\begin{figure*}[tb]
    \centering
    \includegraphics[width=\textwidth]{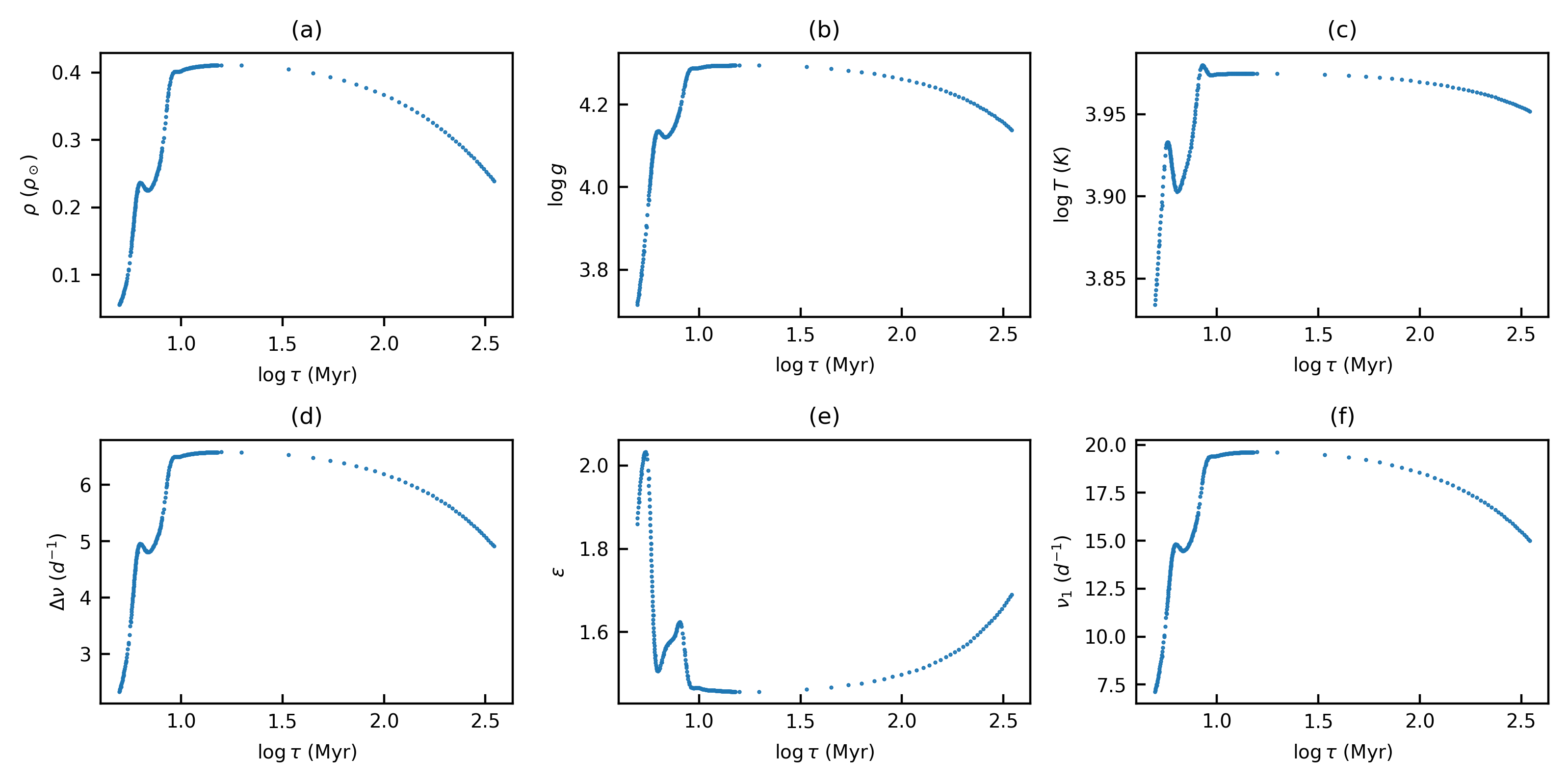}
    %\captionsetup{labelfont={color=blue, bf}}
    \caption{Evolution with time of stellar properties for a model with mass 2.2 $M_\odot$ and $Z = 0.026$. Most structural and seismic quantities cross their pre-MS values during MS. Hence, values of these parameters are degenerate at two different ages. Crossing effects are shown for (a) $\rho$, (b) $\log g$, (c) $\log \Tf$, (d) $\Dn$, (e) $\eps$ and (f) $\nu_1$ (frequency corresponding to $n=1, \ell=0$).}
    \label{fig:t-cross}
\end{figure*}

From Figure~\ref{fig:validation} (c), it is evident that while age inferences are accurate for younger ($\log \tau / \rm{Myr} < 1$) and older stars ($\log \tau / \rm{Myr} > 2.5$), the network systematically over-predicts for the stars of intermediate age ($\log \tau / \rm{Myr} \sim 1.3 - 2.5$). Possible reason for this discrepancy is that no observable evolves as a monotonic function of age. All structure and seismic parameters cross their pre-main-sequence values sometimes within the main sequence, as shown in Figure \ref{fig:t-cross}. Hence, none of our inputs may be used to uniquely distinguish between two possible ages. This fact was also discussed in \citealt{seismic_age}.

We implemented our trained networks on 43 $\delta$ Sct stars from \citet{nature}, for which {$\Tf, L, \Dn, \eps$} measurements were simultaneously available. While investigating the results, we noticed that most of the inferences have unusually high metalicity of $Z$ = 0.026, which is also the upper boundary of our model grid.

It is well-known that what the neural networks learn is not easily interpretable (\citealt{interpretml}). This makes it impossible to investigate and mitigate the shortcomings with the method.

\subsubsection{Feature Importance} \label{method1_feature}
\begin{figure*}[tb]
    \includegraphics[width=\textwidth]{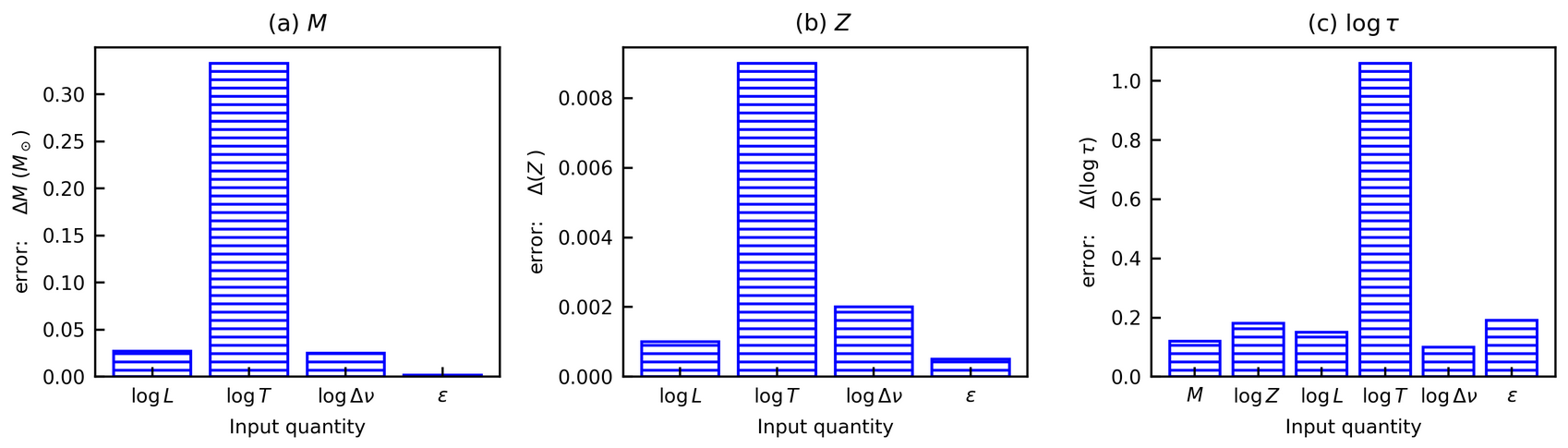}
    %\captionsetup{labelfont={color=blue, bf}}
    \caption{Different inputs contribute unequally towards overall inferences of (a) $M$, (b) $Z$ and (c) $\log \tau$. This figure qualitatively depicts the contribution strengths corresponding to the input parameters when perturbed by 0.5\%.}
    \label{fig:contribution}
\end{figure*}

We determined the qualitative importance of various inputs to determine how strongly they influence parameter estimation. To measure the independent contributions arising from the inputs, we perturbed each input quantity by 0.5\% (without changing other inputs) and measured the relative differences in outputs. We expected the contribution of the input to be proportional to the relative difference in output.
We carried out this process for all input quantities to determine their qualitative importance. Figure \ref{fig:contribution} shows the average relative differences, which can be taken as a proxy for feature importance.

Figure~\ref{fig:contribution} indicates that $\log \Tf$ contributes the most towards the inference of all parameters.  $\Tf$ is strongly influenced by the conditions at the stellar core ($T_c$) through various transport and mixing processes. Hence stellar age, which strongly depends on core hydrogen abundance ($X_c$), can be expected to correlate with $T_c$ and hence, indirectly with $\Tf$.

$M$ and $Z$ are necessary inputs for the inference of age, which is straightforward from the inversion relation discussed above. Despite being an offset parameter that controls the shift in the $\ell=0$ frequencies, $\eps$ plays an important role in the inference of $\tau$. This was also emphasized in \citealt{nature}, where the $\Dn-\eps$ diagram was shown to encode information about age ($\tau$).

Apart from $\Tf$, $M$ significantly depends on $L$ which is likely due to the well known ``mass-luminosity'' relation. Its dependence on $\Dn$ is not surprising owing to the fact that $M$ is related to density ($\rho$) and $\Dn \propto \sqrt{\rho}$.

$\Dn$ plays an important role in the inference of $Z$. It is highly correlated with the square root of mean stellar density -- which is tightly related to metallicity ($Z$) because the higher the metallicity, the lower the stellar density.

\subsection{Method 2: ML using radial modes} \label{method2}
Although we used $L$ as an input quantity in Method-1, the reliability of its measurements depends on the accuracy of stellar distance, as $L$ is calculated using magnitude and distance. Since structure parameters are very sensitive to $\log \Tf$  (evident from section \ref{method1_feature}), errors in measuring the latter propagate into inferences of the former.

In this section, we show that if we use first seven eigenfrequencies of radial-mode oscillations instead of the observables ($L,\Tf, \Dn, \eps$), we still achieve similar results while overcoming existing issues. Reason behind such selection is that (i) all the synthetic models at least seven overtones of radial modes and (ii) radial modes are far more easily identifiable than dipole modes using period ratios (\cite{Petersen1996}) and multicolor photometry (\citealt{multicolor}) like techniques.

Architecture and training of this network is similar to that of Method-1, except its input layer comprising of seven neurons to accept the seven radial modes.

\begin{figure*}[tb]
    \includegraphics[width=\textwidth]{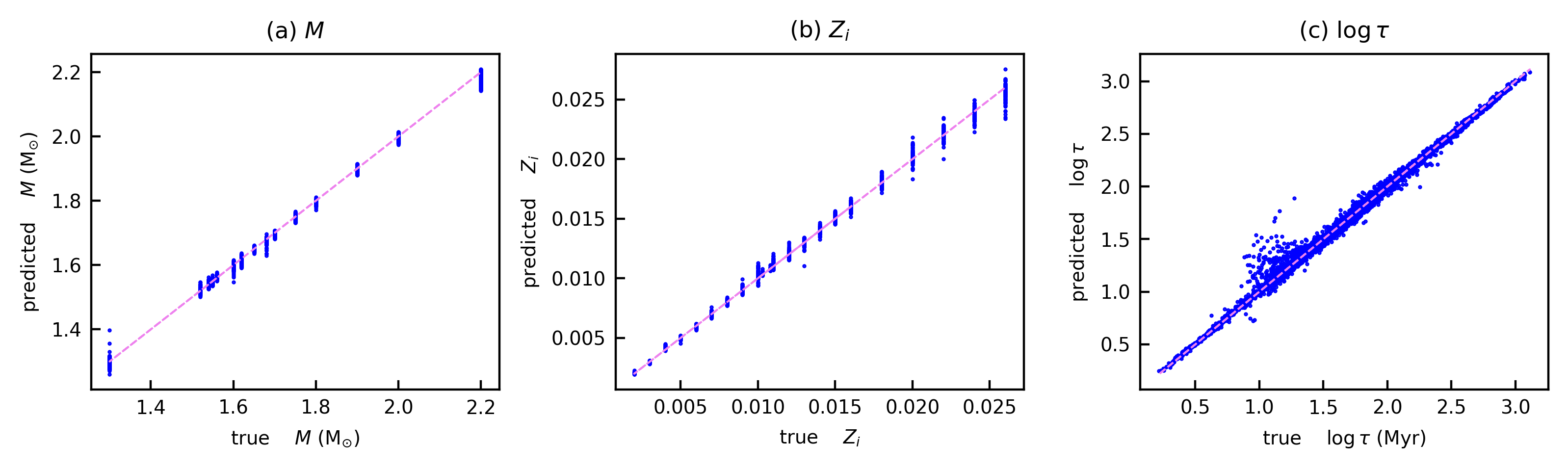}
    %\captionsetup{labelfont={color=blue, bf}}
    \caption{Network predictions for validation samples for (a) $M$, (b) $Z_i$ and (c) $\log\tau$, on taking eigenfrequencies \{$\nu_1, \dots, \nu_7$\} as inputs (without $L$, $\Tf$, $\Dn$, $\eps$). This implies that these non-asymptotic modes are helpful for parametric inference, particularly useful when $L$ or $\Tf$ are not available or reliable.}
    \label{fig:robust}
\end{figure*}

In this formalism, neural networks could learn all the parameters, even without needing \{$L, \Tf, \Dn, \eps$\}, as seen in Figure~\ref{fig:robust}.
We compare the average learning accuracy ( = 100\% - average\% error) associated with each parameter inference using the current method and Method-1 of Section~\ref{method1}. This method is as accurate as Method-1 in inferring $M$ and $Z$. However, it infers age at a significantly higher accuracy (Table \ref{t:earlier-rob-com}). We designate this procedure as \textit{Method-2} in order to distinguish it from Method-1.

\begin{table}[htb]
    %\centering
    \begin{tabular}{ccc}
    \hline \hline
                 & Method-1                & Method-2 \\
                 & (Section \ref{method1}) & (Current Method) \\
         Inputs: & ($L, \Tf, \Dn, \eps$)   & \{$\nu_1 - \nu_7$\} \\ \hline
         $M$     & 99.9\%                 & 99.5\% \\
         $Z$     & 99.7\%                 & 99.5\% \\
         $\log\tau$  & 93\%                  & 96\% \\ \hline
    \end{tabular}
    \caption{Comparing performances as quantified using average learning accuracy of the robust method discussed in this section and Method-1 of Section \ref{method1}. While both perform similarly, Method-2 works without \{$L, \Tf, \Dn, \eps$\}.}
    \label{t:earlier-rob-com}
\end{table}

This method highlights the importance of the $\ell=0$ eigenfrequencies of radial orders $n = 1 - 7$. It shows that the non-asymptotic modes succinctly preserve the representation of $L$ and $\Tf$, although how this is so is unclear.

Despite its robustness, we were unable to apply this method to the observed stars since accurately identifying continuous radial overtones ($\nu_1 - \nu_7$) is difficult. Additionally, it was harder to automate this method for multiple stars because identifying radial orders requires significant human intervention. Identifying radial modes crucially depends on choosing the correct $\Dn$, without which \'echelle diagrams cannot be constructed. Even if this method worked efficiently, we were unable to resolve the age discrepancy (similar as Method-1) over the synthetics, as evident from Figure \ref{fig:robust} (c). In Section~\ref{method3}, we propose another method that improves over this issue.

\subsubsection{Feature Importance} \label{method2_feature}
\begin{figure*}[tb]
    \centering
    \includegraphics[width=\textwidth]{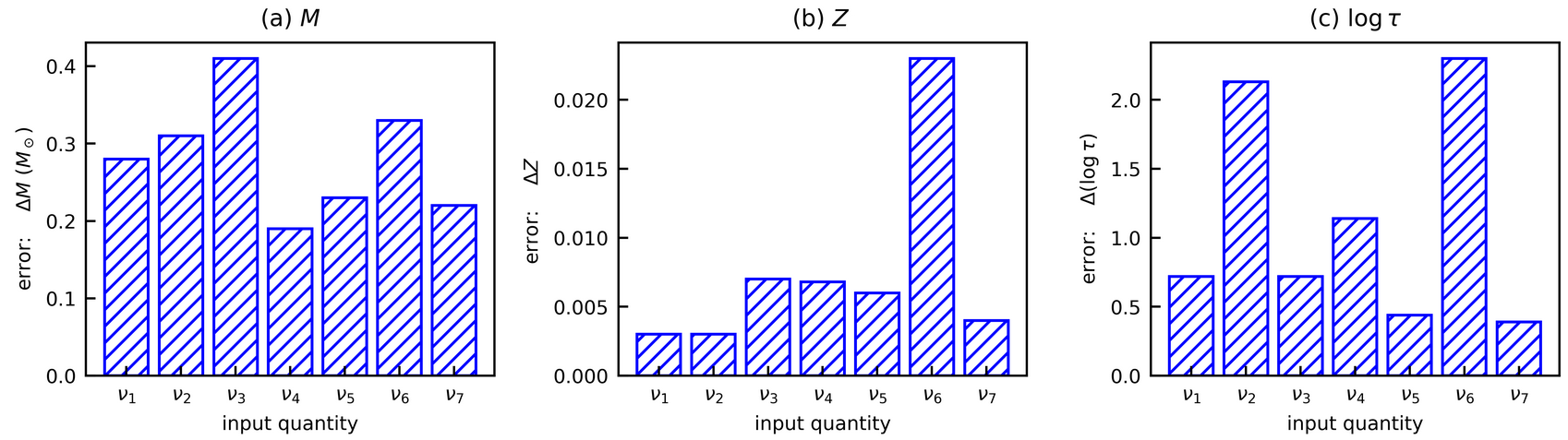}
    %\captionsetup{labelfont={color=blue, bf}}
    \caption{Importance of first 7 eigenfrequencies in the inference of (a) $M$, (b) $Z$ and (c) $\log \tau$.}
    \label{fig:robust-feature}
\end{figure*}

%%%%%%%%%%%%%%%%%%%%%%%%%%%%%%%%%%%%%%%%%%%%%%%%%%%%%%%%%%%
We investigated the strengths of contributions arising from each mode involved in the inference routine. We measured their importance in a manner similar to that in Section \ref{method1_feature}. The attendant measures are shown in Figure~\ref{fig:robust-feature}. It may be understood from the figure that all frequencies are not equally important. $\nu_6$ has the highest contribution, followed by $\nu_2$ and $\nu_3$. Hence, both non-asymptotic and asymptotic modes seem to be important.

It is surprising that the network depends so significantly on $\nu_6$, robust over a range of different $\Dn$ and $\eps$. The probable cause is that the transition to asymptotic nature ($n \gg \ell$) from low-order non-asymptotic modes occurs around $\nu_6$. It is possible that $\nu_6$ therefore contains information from both regimes.

Having emphasized the salient features of $\nu_6$, it is interesting that the network does not equally depend on modes of radial order 4, 5 and 7 - which altogether constitute the asymptotic series and hence carry similar $\Dn$ characteristics. This may be similar to principal component analysis (PCA) or dimensionality reduction algorithms, and the network is trying to rely on as few inputs as possible, leaving others redundant.

\subsection{Method 3: ML using radial and dipole modes} \label{method3}
In this section, we present the final follow-up experiment to study the importance of dipole modes. Every experiment carried out so far has dealt with radial ($\ell = 0$) modes, or quantities such as \{$\Dn,\eps$\} that depend on radial modes. However, since dipole modes are often seen in the \'echelle diagrams (\citealt{nature}) of $\delta$ Sct stars, it is useful to conduct these experiments in order to study their contribution.

In this method, we built a model that took observables \{$L, \Tf$\} and eigenfrequencies with $n \in [4-7]$ from each of the $\ell=$ 0 and 1 ridges, and produced as output the structure parameters. For age inference, we supplemented pre-inferred $M$ and $Z$ to the frequencies and \{$L, \Tf$\}. We refer to this procedure as \textit{Method-3}.

\begin{figure*}[tb]
    \includegraphics[width=\textwidth]{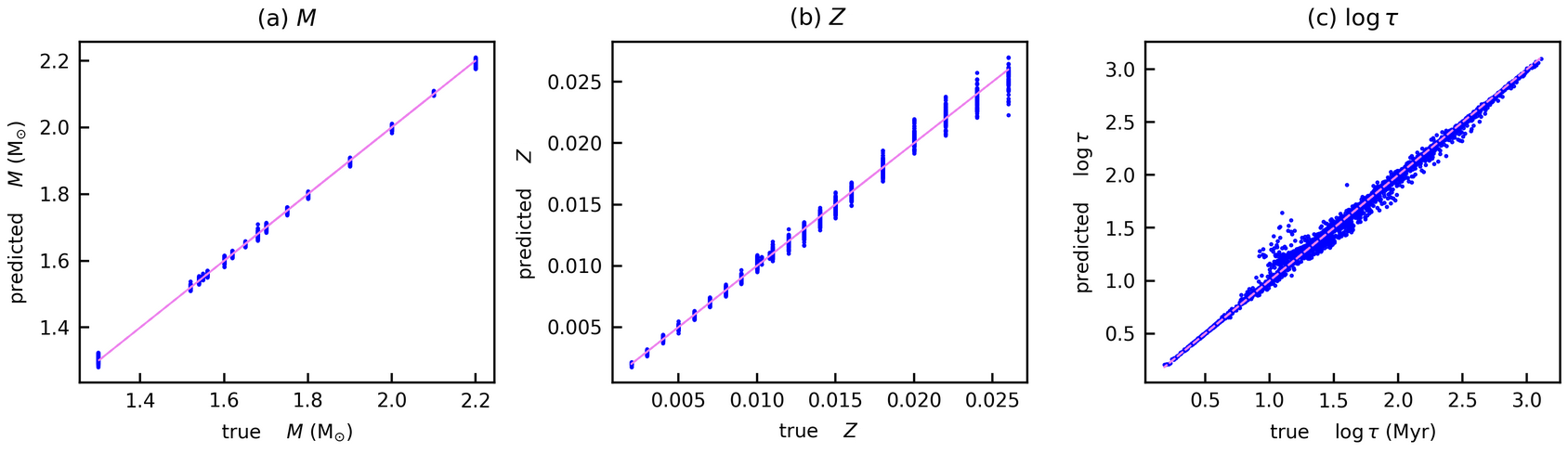}
    %\captionsetup{labelfont={color=blue, bf}}
    \caption{Improved network prediction after inputting radial-dipole ($\ell = 0, 1$) modes of overtone $n \in [4 - 7]$ along with \{$L, \Tf$\}. The age inference is much more accurate than the approaches described earlier. \label{improving-age-problem}}
\end{figure*}

Figure~\ref{improving-age-problem} and Table~\ref{t:earlier-rob-dipole-com} indicate that the networks were able to learn $M$,$Z$, and $\log \tau$ from synthetics and the accuracy of age inference has improved.
Since the current method adequately constrains age, this emphasizes the importance of dipole modes.

\begin{table}[htb]
    \centering
    \begin{tabular}{cccc}
    \hline \hline
                 & Method-1                 & Method-2                   & Method-3 \\
                 & (Section \ref{method1})  & (Section \ref{method2})    & (Current Method) \\
         Inputs: & \{$L, \Tf, \Dn, \eps$\}  & \{$\nu_1 - \nu_7$\}        & \{$L, \Tf,$ 8 modes\} \\ \hline
         $M$     & 99.9\%                  & 99.5\%                    & 99.8\% \\
         $Z$     & 99.7\%                  & 99.5\%                    & 99.6\% \\
         $\log\tau$  & 93\%                & 96\%                      & 98.1\% \\ \hline
    \end{tabular}
    \caption{Comparing performances (average learning accuracy) of Methods-1 and 2 (Section \ref{method1}, \ref{method2}) with Method-3 (this section), where we use $\Tf, L$, 4 radial and 4 zonal dipole modes as inputs.}
    \label{t:earlier-rob-dipole-com}
\end{table}

With more number of inputs to neural network, it is supposed to gain higher accuracy. We also measured $\chi^2/N$ for all the three experiments to assess which method is more efficient. Here $\chi^2$ mean average squared differences between true and machine predicted values and $N$ represents the number of inputs fed to the networks. Table \ref{t:chi_sq_div_N_method123} summarizes these facts.

\begin{table}[htb]
    \centering
    \begin{tabular}{cccc}
    \hline \hline
                 & Method-1                 & Method-2                   & Method-3 \\
                 & (Section \ref{method1})  & (Section \ref{method2})    & (Current Method) \\
         Inputs: & \{$L, \Tf, \Dn, \eps$\}  & \{$\nu_1 - \nu_7$\}        & \{$L, \Tf,$ 8 modes\} \\ \hline
         $M$     & $7\times10^{-5}$         & $10^{-5}$                  & $2\times10^{-6}$ \\
         $Z$   & $10^{-7}$                & $5\times10^{-8}$           & $9\times10^{-9}$ \\
         $\log\tau$  & $3\times10^{-3}$     & $6.5\times10^{-4}$         & $10^{-3}$ \\ \hline
    \end{tabular}
    \caption{Comparing $\chi^2/N$ for all the methods.}
    \label{t:chi_sq_div_N_method123}
\end{table}

We observe that Method-2 consistently outperforms Method-1. Method-3 excels among all of these but for age inference it could not outperform the Method-2. This may be due to the relatively larger (yet tolerable) spread at larger ages compared to Method-2. Global statistics such as the $\chi^2$ average out-small scale information and hence cannot be used to assess the performance across all the ages. However a performance-wise comparison after studying Figures \ref{fig:robust} and \ref{improving-age-problem} suggest that the outlier level is lower for age inference in Method-3. Hence, we conclude that results from Method-3 are preferred. Nonetheless, Method-1 remains practically applicable until definitive mode identification becomes possible in $\delta$ Sct stars.

At present, we were unable to apply this method to measure structure parameters in observed stars as it is challenging to accurately label the dipole modes.

\subsubsection{Feature Importance} \label{method3_feature}
\begin{figure*}[tb]
    \centering
    \includegraphics[width=\textwidth]{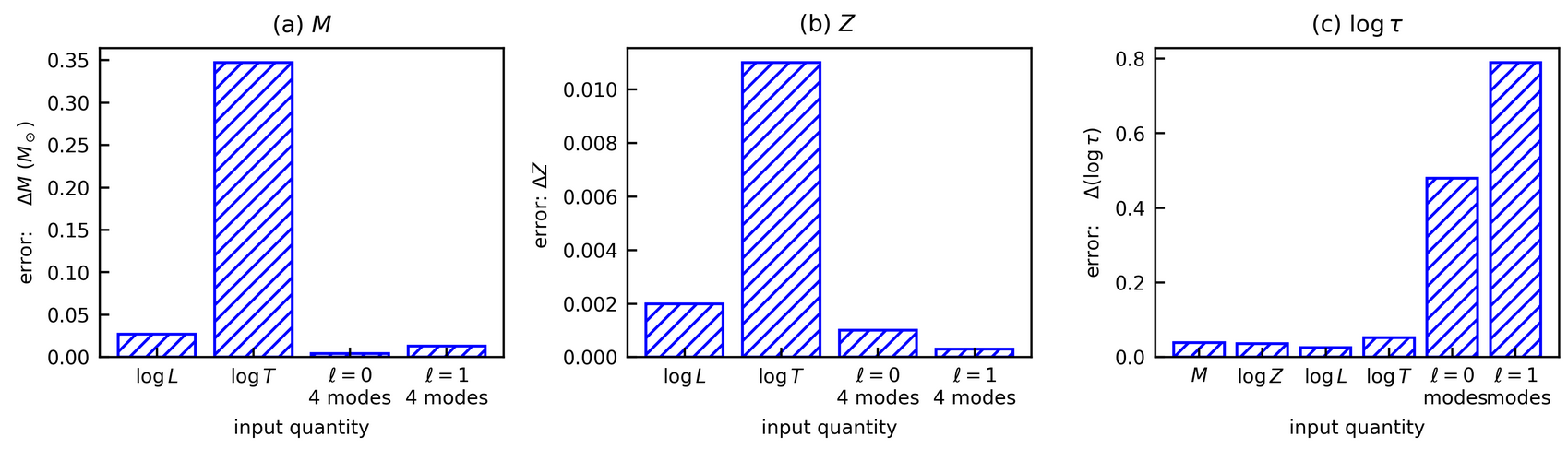}
    \caption{Qualitative contributions arising from input quantities for inferences of (a) $M$, (b) $Z$ and (c) $\log\tau$.}
    \label{fig:col-robust-contribution}
\end{figure*}

%%%%%%%%%%%%%%%%%%%%%%%%%%%%%%%%%%%%%%%%%%%%%%%%%%%%%%%%%%%
To understand the importance of the dipole and radial modes, we carried out a feature-importance experiment for these parameters and show the plots in Figure \ref{fig:col-robust-contribution}.

An important conclusion is that while $M$ and $Z$ do not significantly depend on radial and dipole frequencies, the exact opposite is seen when inferring age. Due to their global nature, $L$ and $\Tf$ appear to encode enough information of global parameters $M$ and $Z$, rendering  the oscillation measurements less useful. In contrast, age depends on the deep interior more than the shallow layers since the core hydrogen fraction ($X_c$) governs the evolution. Since pulsation frequencies are ideal probes of the stellar interior, age determination strongly relies on them.

%\subsection{Method 1: ML using seismic indices} \label{method1}
%\input{method-1}

%\subsection{Method 2: ML using radial modes} \label{method2}
%\input{method-2}

%\subsection{Method 3: ML using radial and dipole modes} \label{method3}
%\input{method-3}

\section{Method 4: Least square fitting of stellar parameters} \label{method4}
Each method described above has its own set of advantages and disadvantages.
In this section, we employed a grid-search algorithm to minimize the least-square loss functions in order to identify the model best matching to the observation. This method requires interactive inspection to obtain accurate fits, provided that there exists a close fit model to the observation.  A similar approach was used by \citealt{Steindl2022} to fit stellar parameters to a small number of $\delta$ Sct stars.

One of the advantages of this method is it can be applied to the 60 stars from \citealt{nature} to achieve good results. Even for stars with missing values of $L$ or $\Tf$, this method can obtain a good fit to observed spectra.

\subsection{Method}
%%%%%%%%%%%%%%%%%%%%%%%
The method of least squares is a common technique to fit photometric ($L$), spectroscopic ($\Tf$) and seismic quantities (eigen-frequencies) between observations and models. The underlying principle involves assigning a $\chi^2$ value (as shown in equation \ref{eq:chisq}) to each sample in the model, which represents a weighted combination of the squared differences between the observed and model values. The objective is to find the model sample that results in the minimum value of $\chi^2$. The values of $\sigma_L$, $\sigma_{T_{\rm eff}}$, and $\sigma_{\nu_i}$ are critical parameters in this approach, as they represent the magnitude of the errors that can be tolerated. This methodology can be highly effective if the appropriate $\sigma$ values are used.

\begin{align} \label{eq:chisq}
\chi^2 =    &  \left\{ \nonumber \dfrac{(L_{\rm obs} - L_{\rm model})^2}{\sigma_L^2} + \dfrac{(T_{\rm eff; obs} - T_{\rm eff; model})^2}{\sigma_{T_{\rm eff}}^2} \right. \\
            & \left. + \sum_{i=1}^{N} \dfrac{(\nu_{i,{\rm obs}}-\nu_{i,{\rm model}})^2}{\sigma_{\nu_i}^2} \right\}   
\end{align}

The last part of equation \ref{eq:chisq} indicates the requirement of determining both the observed ($\nu_{i,{\rm obs}}$) and model frequencies ($\nu_{i,{\rm model}}$) of the same radial order and angular degree. However, accurately identifying and labeling the observed modes in $\delta$ Sct stars is challenging due to the high density contamination, missing modes as well as presence of modes with unknown origin. Identification of genuine modes requires iterative pre-whitening followed by elimination of frequency combinations -- which adds to the complexity of the fitting routine.

In our approach, we have made slight modifications to a similar method. Our primary objective is to search for a model that can reprodcue as many observed modes as possible without having additional spurious peaks.
We have applied constraints to the models using the observed values of $\Dn$ (which are obtained from \citealt{nature}) to prevent selection of erroneous fits - since a model with much lower $\Dn$ can accurately match a relatively large number of significant observed modes.
Finally, we introduce a $\chi^2$ term to compare few observed frequencies to their nearest model frequencies ($\nu^{\rm closest}_{\rm model}$). Our modified $\chi^2$ looks like equation \ref{eq:chisq_new}.

\begin{align} \label{eq:chisq_new}
\chi^2  = & \left\{ \nonumber \dfrac{(L_{\rm obs} - L_{\rm model})^2}{\sigma_L^2}  \right. \\
        & \nonumber + \dfrac{(T_{\rm eff; ~obs} - T_{\rm eff; ~model})^2}{\sigma^2_{T_{\rm eff}}} \\
        & \nonumber + \dfrac{(\Dn_{\rm obs} - \Dn_{\rm model})^2}{\sigma^2_{\Delta\nu}} \\
        & \left.+ \sum_{\nu_1}^{\nu_N} \dfrac{(\nu_{\rm obs}-\nu^{\rm closest}_{\rm model})^2 }{\sigma^2_\nu} \right\}
\end{align}
The $\chi^2$ formula therefore has similarity to that of \cite{Steindl2022} and the likelihood function of \cite{scutt2023asteroseismology}. 

Values of $\sigma_L$ have been taken from \cite{nature}. Following the same, we have set the uncertainty in $T_{\rm eff}$ to be 2\% and that of $\Delta\nu$ to be 0.02 ${\rm d^{-1}}$. Rayleigh frequency resolution criterion has been taken as the uncertainty in mode frequencies, similar to \cite{Steindl2022}. Since we fitted three parameters simultaneously ($M, Z, \tau$), $1\sigma$ uncertainties (or 68\% confidence interval) associated with the best-fit parameters correspond to 3.5 increase from $\chi^2_{\rm min}$ (Table 1 of \cite{Avni}). Hence we assembled all the model parameters corresponding to $\chi^2 \in [\chi^2_{\rm min}, \chi^2_{\rm min}+3.5]$ and used their spread to calculate the uncertainties.

Regarding the frequency component of the $\chi^2$, we neither calculated eigen-frequenies from iterative pre-whitening nor assigned any possible identifications to the peaks. Instead, we visually inspected the spectrum and started by selecting four peaks of relatively high frequency and significant amplitudes. We selected these modes through trial and error. After begining with these arbitrary set of modes we searched for their closest eigen-frequencies ($\nu^{\rm closest}_{\rm model}$) across all the model samples disregarding their $n$ or $\ell$. Treating these frequenies as inputs to equation \ref{eq:chisq_new}, we calculated $\chi^2$ across all the models and inspected for $\chi^2_{\rm min}$. We repeatedly selected different sets of input frequencies as well as varied the number of fitted modes (N = 1, 2, 3, 4, rarely 5) and carried out the entire fitting process afresh until we achieved a minimum possible $\chi^2_{\rm min}$. Finally, we also constructed an \`echelle diagram to ensure that the $N$ selected modes were genuine $m=0$ frequencies and the obtained solution was able to fit other significant radial and dipole modes as well. Hence, we used the \`echelle diagrams to confirm that we are not comparing model modes with incorrect peaks such as rotational splits, combination frequencies etc.

The reason we favoured selecting very few modes is as follows. It is not always possible to obtain exact fits to all the observed modes simultaneously. In such scenarios, fitting larger numbers of modes would demand a highly sophisticated mode selection process since we are picking up the modellable modes through trial and error. Inappropriate mode selection usually leads to a solution where none of the modes are comparable to the observed modes. However, for each star, we were successful to identify a smaller number of genuine modes using which we could fit most of the observed modes.

This fitting routine takes us $\sim$ 10 seconds given that we vectorized this operation across 112 cpu cores using NumPy. Otherwise, this would correspond to $\sim$ 20 cpu minutes (per single star) without any core-level parallelization. However the mode selection process (being manual and interactive) is the most difficult one and it took us around 10 minutes per a single star.

\begin{figure*}[tb]
    \includegraphics[width=\textwidth]{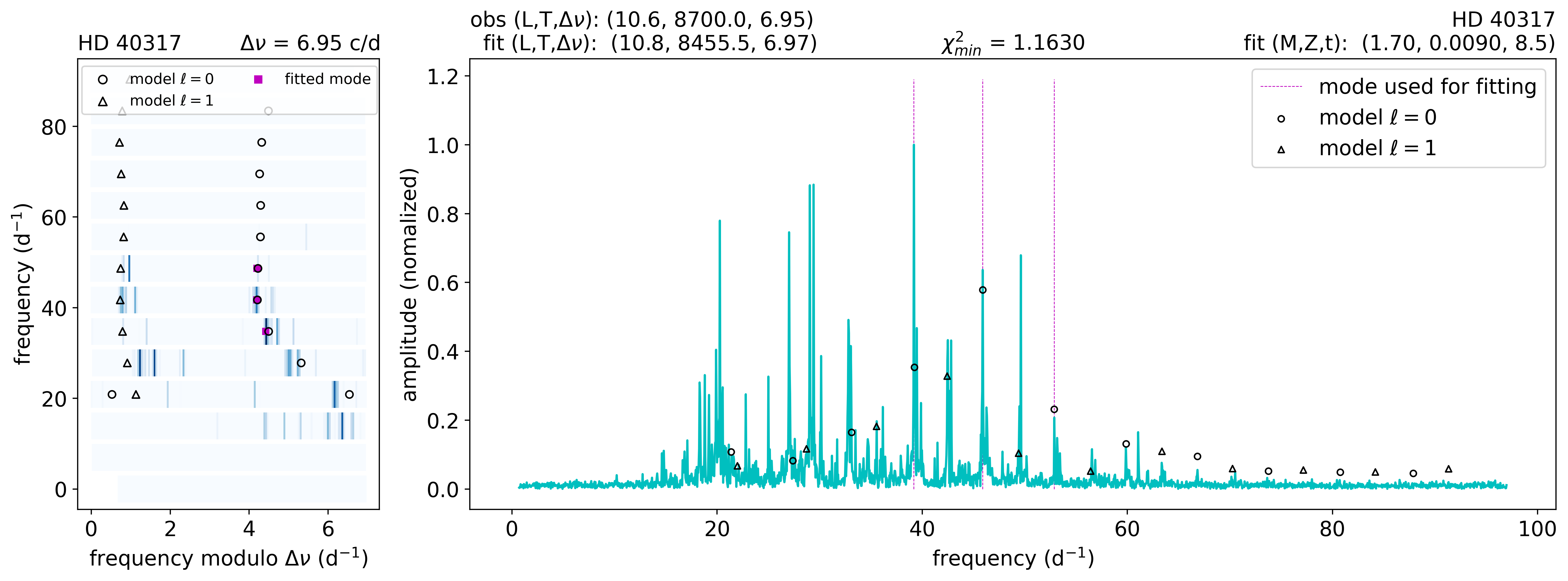}
    %\captionsetup{labelfont={color=blue, bf}}
    \caption{ Obtaining a best-fit model to observation of HD 40317. The dotted lines in the right side spectrum shows the 3 modes we selected for the fitting routine. These 3 modes are also shown in the \'echelle diagram (on left) as filled rectangle symbols. After determining the best-fit model, we show its radial and dipole modes as open circles and triangles over both the \'echelle diagram and the oscillation spectrum. Some of these open symbols do not exactly fall on the observed modes because the fitted frequencies fall within the spectral windows in the vicinity of the sharp peaks. We also report the observed and best-fitted values of $L, \Tf$ and $\Dn$ in the title of the spectrum plot. The ($M, Z, \tau$) values of the best fitted model are presented in the same title. \label{fig:close-fit-example}}
\end{figure*}

In Figure \ref{fig:close-fit-example}, we present an example of our successfully fitted results using this methodology. It was challenging for us to find a close-fit model where lower-order model modes would precisely align with the observed modes. Additionally, dipole modes of the fitted models were sometimes seen to be present in the vicinity of the observed ones but not precisely. This discrepancy could possibly be due to the unequal splitting of dipole modes and the rotation-induced shift of $m=0$ components, even in radial modes.

\subsection{Results}
%%%%%%%%%%%%%%%%%%%%%%%

In this section, we present the results of our method on 60 stars from \citealt{nature}. For 3 of these stars, either one or both of ($L, \Tf$) inputs were missing. For such stars, we ignored the corresponding contributions of $\chi^2$. As the best fit model has an inherent value of missing parameters, one can have a crude estimate of the parameters for these stars.

\begin{figure*}[tb]
    \includegraphics[width=\textwidth]{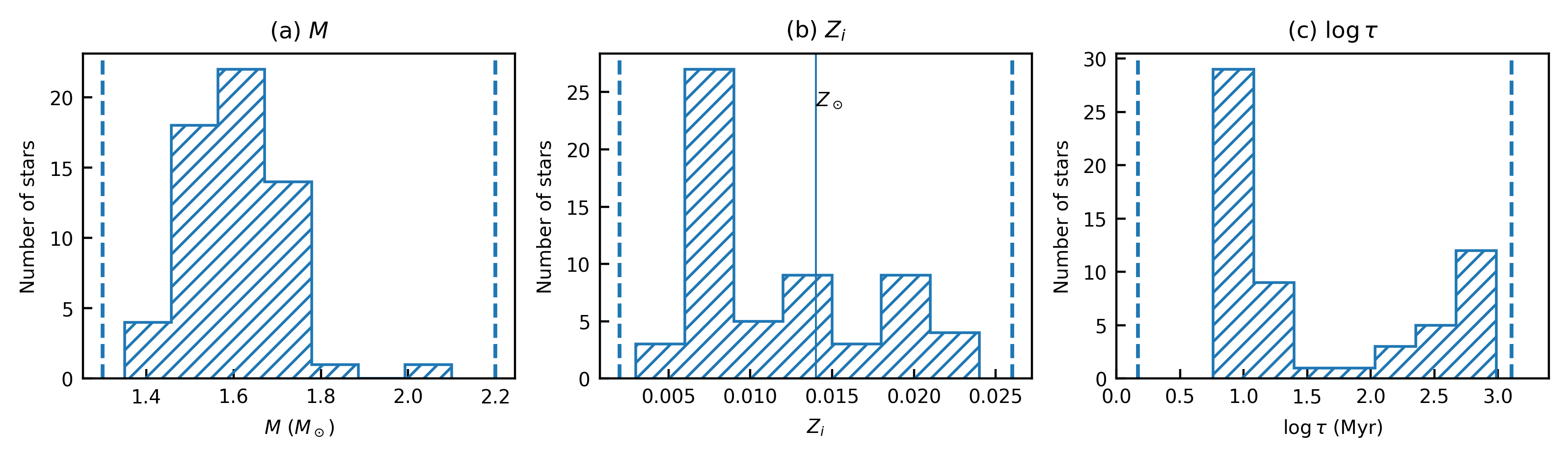}
    %\captionsetup{labelfont={color=blue, bf}}
    \caption{Statistics of the fundamental parameters of 60 $\delta$ Sct stars, inferred using the least-squares minimization method. The dotted vertical lines represent the boundary of the model grid.}
    \label{real-hist}
\end{figure*}

In Table \ref{table-result} and Figure \ref{inf-real-M} - \ref{inf-real-t} of appendix \ref{inf-real}, we have presented the fitted structure parameters of the individual stars. Additionally, Figure \ref{real-hist} summarizes the statistics, indicating that most of the fitted masses are approximately $\sim 1.6 ~\text{M}_\odot$, while the dominant metallicity values are distributed around $Z = 0.010$, corresponding to [Fe/H] $\sim -0.146$. We also observed a bi-modal age distribution among the stars, where more than half of them are young (around $\sim 10$ Myr) and a few are very old (over $100$ Myr). However, we emphasize that age inferences are not highly precise due to the degeneracy effect, which means that at two different ages, the star can have similar physical and seismic structure.

\startlongtable
\begin{deluxetable}{cDDD}
\tablecaption{Inferred values of mass, composition, and age of 60 $\delta$ Sct stars taken from \cite{nature}. These are obtained from the $\chi^2$ minimization method.}
\label{table-result}
\tablehead{
\colhead{Id} & \multicolumn2c{mass}        & \multicolumn2c{$Z_i$} & \multicolumn2c{age} \\
\colhead{}   & \multicolumn2c{($M_\odot$)} & \multicolumn2c{}      & \multicolumn2c{(Myr)}
}
\decimals
\startdata
%\\
HD 2280            & 1.35 $^{+0.1 } _{-0.05}$ & 0.008  $^{+0.001 } _{-0.002 }$ & 977.7 $^{+202.9} _{-  1.23 }$ \\ \vspace{0.24cm}
HD 3622            & 1.7  $^{+0.05} _{-0.15}$ & 0.022  $^{+0.002 } _{-0.014 }$ & 50.6  $^{+ 75.3 } _{- 42.04}$ \\ \vspace{0.24cm}
HD 10779           & 1.6  $^{+0.1 } _{-0.02}$ & 0.009  $^{+0.013 } _{-0.001 }$ & 9.0   $^{+156.43} _{-  0.25 }$ \\ \vspace{0.24cm}
HD 17341           & 1.65 $^{+0.1 } _{-0.01}$ & 0.012  $^{+0.006 } _{-0.001 }$ & 9.0   $^{+  2.67 } _{-  0.01 }$ \\ \vspace{0.24cm}
HD 17693           & 1.7  $^{+0.05} _{-0.05}$ & 0.012  $^{+0.001 } _{-0.003 }$ & 9.7   $^{+  0.01 } _{-  1.59 }$ \\ \vspace{0.24cm}
HD 20203           & 1.5  $^{+0.04} _{-0.05}$ & 0.008  $^{+0.001 } _{-0.001 }$ & 692.5 $^{+  0.06 } _{-132.78}$ \\ \vspace{0.24cm}
HD 20232           & 1.68 $^{+0.02} _{-0.13}$ & 0.018  $^{+0.002 } _{-0.011 }$ & 229.8 $^{+104.08} _{-221.47}$ \\ \vspace{0.24cm}
HD 24572           & 1.52 $^{+0.06} _{-0.02}$ & 0.013  $^{+0.003 } _{-0.004 }$ & 344.0 $^{+132.62} _{-331.75}$ \\ \vspace{0.24cm}
HD 24975           & 1.7  $^{+0.05} _{-0.06}$ & 0.014  $^{+0.001 } _{-0.003 }$ & 10.4  $^{+  0.6  } _{-  1.55 }$ \\ \vspace{0.24cm}
HD 25674           & 1.54 $^{+0.01} _{-0.04}$ & 0.007  $^{+0.001 } _{-0.001 }$ & 829.4 $^{+ 77.82} _{-  0.08 }$ \\ \vspace{0.24cm}
HD 28548           & 1.7  $^{+0.05} _{-0.06}$ & 0.012  $^{+0.001 } _{-0.006 }$ & 163.0 $^{+158.0} _{-155.63}$ \\ \vspace{0.24cm}
HD 30422           & 1.5  $^{+0.02} _{-0.05}$ & 0.006  $^{+0.001 } _{-0.001 }$ & 8.3   $^{+  0.01 } _{-  0.01 }$ \\ \vspace{0.24cm}
HD 31322           & 1.75 $^{+0.05} _{-0.05}$ & 0.011  $^{+0.001 } _{-0.002 }$ & 7.7   $^{+  0.01 } _{-  0.75 }$ \\ \vspace{0.24cm}
HD 31640           & 1.55 $^{+0.09} _{-0.03}$ & 0.009  $^{+0.004 } _{-0.001 }$ & 9.3   $^{+  2.44 } _{-  0.23 }$ \\ \vspace{0.24cm}
HD 31901           & 1.58 $^{+0.06} _{-0.02}$ & 0.009  $^{+0.001 } _{-0.001 }$ & 9.4   $^{+  0.01 } _{-  0.58 }$ \\ \vspace{0.24cm}
HD 32433           & 1.62 $^{+0.03} _{-0.1 }$ & 0.018  $^{+0.002 } _{-0.01  }$ & 201.1 $^{+193.23} _{-191.66}$ \\ \vspace{0.24cm}
HD 38597           & 1.58 $^{+0.06} _{-0.02}$ & 0.008  $^{+0.002 } _{-0.001 }$ & 694.8 $^{+  0.02 } _{-185.18}$ \\ \vspace{0.24cm}
HD 38629           & 1.45 $^{+0.05} _{-0.1 }$ & 0.004  $^{+0.001 } _{-0.001 }$ & 972.4 $^{+  0.21 } _{-  0.86 }$ \\ \vspace{0.24cm}
HD 40317           & 1.7  $^{+0.05} _{-0.08}$ & 0.009  $^{+0.001 } _{-0.002 }$ & 8.5   $^{+  0.16 } _{-  1.17 }$ \\ \vspace{0.24cm}
HD 42005           & 1.6  $^{+0.02} _{-0.02}$ & 0.008  $^{+0.001 } _{-0.001 }$ & 8.6   $^{+  0.01 } _{-  0.01 }$ \\ \vspace{0.24cm}
HD 42608           & 1.75 $^{+0.05} _{-0.1 }$ & 0.018  $^{+0.002 } _{-0.01  }$ & 13.8  $^{+ 96.09} _{-  5.7  }$ \\ \vspace{0.24cm}
HD 44726           & 1.55 $^{+0.01} _{-0.03}$ & 0.011  $^{+0.001 } _{-0.003 }$ & 396.6 $^{+253.75} _{-383.59}$ \\ \vspace{0.24cm}
HD 44930           & 1.64 $^{+0.01} _{-0.1 }$ & 0.018  $^{+0.008 } _{-0.005 }$ & 13.7  $^{+453.78} _{-  3.78 }$ \\ \vspace{0.24cm}
HD 44958           & 1.6  $^{+0.1 } _{-0.02}$ & 0.009  $^{+0.013 } _{-0.001 }$ & 9.1   $^{+ 11.22} _{-  0.05 }$ \\ \vspace{0.24cm}
HD 45424           & 1.55 $^{+0.01} _{-0.03}$ & 0.008  $^{+0.001 } _{-0.001 }$ & 823.0 $^{+ 54.73} _{-  0.06 }$ \\ \vspace{0.24cm}
HD 46722           & 1.52 $^{+0.01} _{-0.02}$ & 0.008  $^{+0.001 } _{-0.001 }$ & 9.1   $^{+  0.27 } _{-  0.01 }$ \\ \vspace{0.24cm}
HD 48985           & 1.64 $^{+0.04} _{-0.02}$ & 0.008  $^{+0.001 } _{-0.001 }$ & 8.5   $^{+  0.01 } _{-  0.67 }$ \\ \vspace{0.24cm}
HD 50153           & 1.7  $^{+0.05} _{-0.02}$ & 0.02   $^{+0.002 } _{-0.002 }$ & 15.9  $^{+  0.01 } _{-  0.01 }$ \\ \vspace{0.24cm}
HD 54711           & 1.62 $^{+0.06} _{-0.12}$ & 0.009  $^{+0.006 } _{-0.003 }$ & 9.8   $^{+378.22} _{-  1.95 }$ \\ \vspace{0.24cm}
HD 55863           & 1.64 $^{+0.01} _{-0.06}$ & 0.009  $^{+0.001 } _{-0.001 }$ & 8.6   $^{+  0.69 } _{-  0.09 }$ \\ \vspace{0.24cm}
HD 59104           & 1.56 $^{+0.02} _{-0.02}$ & 0.018  $^{+0.002 } _{-0.002 }$ & 19.1  $^{+297.27} _{-  1.96 }$ \\ \vspace{0.24cm}
HD 59594           & 1.6  $^{+0.05} _{-0.1 }$ & 0.015  $^{+0.005 } _{-0.007 }$ & 13.9  $^{+476.0} _{-  4.77 }$ \\ \vspace{0.24cm}
HD 78198           & 1.62 $^{+0.06} _{-0.02}$ & 0.014  $^{+0.006 } _{-0.001 }$ & 9.9   $^{+  4.74 } _{-  0.37 }$ \\ \vspace{0.24cm}
HD 99506           & 1.65 $^{+0.03} _{-0.1 }$ & 0.018  $^{+0.002 } _{-0.01  }$ & 16.2  $^{+246.96} _{-  7.1  }$ \\ \vspace{0.24cm}
HD 223011          & 1.62 $^{+0.18} _{-0.02}$ & 0.01   $^{+0.014 } _{-0.001 }$ & 8.7   $^{+438.49} _{-  0.32 }$ \\ \vspace{0.24cm}
HD 290799          & 1.55 $^{+0.01} _{-0.01}$ & 0.004  $^{+0.001 } _{-0.001 }$ & 7.1   $^{+  0.01 } _{-  0.93 }$ \\ \vspace{0.24cm}
TIC 349645354      & 1.75 $^{+0.05} _{-0.05}$ & 0.024  $^{+0.002 } _{-0.004 }$ & 12.3  $^{+  0.01 } _{-  2.14 }$ \\ \vspace{0.24cm}
TIC 431695696      & 1.54 $^{+0.08} _{-0.04}$ & 0.006  $^{+0.002 } _{-0.001 }$ & 8.1   $^{+  1.14 } _{-  0.01 }$ \\ \vspace{0.24cm}
TIC 124381332      & 1.68 $^{+0.07} _{-0.06}$ & 0.014  $^{+0.001 } _{-0.007 }$ & 13.3  $^{+145.36} _{-  5.31 }$ \\ \vspace{0.24cm}
TIC 340358522      & 1.62 $^{+0.02} _{-0.02}$ & 0.008  $^{+0.001 } _{-0.001 }$ & 8.4   $^{+  0.01 } _{-  0.01 }$ \\ \vspace{0.24cm}
HD 187547          & 1.58 $^{+0.02} _{-0.04}$ & 0.018  $^{+0.002 } _{-0.002 }$ & 177.9 $^{+  0.64 } _{-160.44}$ \\ \vspace{0.24cm}
KIC 8415752        & 1.75 $^{+0.05} _{-0.05}$ & 0.024  $^{+0.002 } _{-0.002 }$ & 344.6 $^{+  0.04 } _{-  0.02 }$ \\ \vspace{0.24cm}
KIC 9450940        & 1.62 $^{+0.03} _{-0.02}$ & 0.015  $^{+0.001 } _{-0.001 }$ & 702.5 $^{+ 53.63} _{- 51.24}$ \\ \vspace{0.24cm}
HD 37286           & 1.52 $^{+0.06} _{-0.02}$ & 0.006  $^{+0.005 } _{-0.001 }$ & 8.5   $^{+  4.51 } _{-  0.01 }$ \\ \vspace{0.24cm}
HD 39060           & 1.45 $^{+0.05} _{-0.1 }$ & 0.008  $^{+0.001 } _{-0.001 }$ & 861.4 $^{+  0.45 } _{-  0.53 }$ \\ \vspace{0.24cm}
HD 42915           & 1.55 $^{+0.09} _{-0.01}$ & 0.006  $^{+0.001 } _{-0.001 }$ & 7.6   $^{+  0.01 } _{-  0.87 }$ \\ \vspace{0.24cm}
HD 290750          & 2.1  $^{+0.1 } _{-0.1 }$ & 0.013  $^{+0.013 } _{-0.001 }$ & 5.8   $^{+185.38} _{-  0.07 }$ \\ \vspace{0.24cm}
TIC 143381070      & 1.5  $^{+0.02} _{-0.05}$ & 0.007  $^{+0.001 } _{-0.001 }$ & 594.4 $^{+  0.03 } _{-117.6}$ \\ \vspace{0.24cm}
TIC 260161111      & 1.54 $^{+0.06} _{-0.04}$ & 0.008  $^{+0.002 } _{-0.003 }$ & 599.3 $^{+ 51.98} _{-591.2}$ \\ \vspace{0.24cm}
HD 10961           & 1.35 $^{+0.1 } _{-0.05}$ & 0.003  $^{+0.001 } _{-0.001 }$ & 8.4   $^{+  0.01 } _{-  0.01 }$ \\ \vspace{0.24cm}
HD 25248           & 1.5  $^{+0.02} _{-0.05}$ & 0.009  $^{+0.005 } _{-0.001 }$ & 10.9  $^{+  4.14 } _{-  0.01 }$ \\ \vspace{0.24cm}
HD 67688           & 1.68 $^{+0.02} _{-0.03}$ & 0.008  $^{+0.001 } _{-0.001 }$ & 7.6   $^{+  0.01 } _{-  0.01 }$ \\ \vspace{0.24cm}
HD 70510           & 1.65 $^{+0.03} _{-0.01}$ & 0.018  $^{+0.002 } _{-0.002 }$ & 90.0  $^{+  0.12 } _{- 74.06}$ \\ \vspace{0.24cm}
HD 75040           & 1.6  $^{+0.02} _{-0.05}$ & 0.011  $^{+0.001 } _{-0.001 }$ & 10.0  $^{+  0.58 } _{-  0.09 }$ \\ \vspace{0.24cm}
HD 222496          & 1.8  $^{+0.05} _{-0.15}$ & 0.022  $^{+0.002 } _{-0.009 }$ & 12.1  $^{+485.56} _{-  3.64 }$ \\ \vspace{0.24cm}
HD 34282           & 1.6  $^{+0.02} _{-0.1 }$ & 0.009  $^{+0.001 } _{-0.004 }$ & 421.4 $^{+390.4} _{-  0.02 }$ \\ \vspace{0.24cm}
HD 29783           & 1.68 $^{+0.02} _{-0.04}$ & 0.009  $^{+0.001 } _{-0.001 }$ & 8.5   $^{+  0.3  } _{-  0.61 }$ \\ \vspace{0.24cm}
HD 220811          & 1.54 $^{+0.04} _{-0.01}$ & 0.015  $^{+0.003 } _{-0.001 }$ & 849.5 $^{+  0.05 } _{-154.81}$ \\ \vspace{0.24cm}
HD 25369           & 1.58 $^{+0.06} _{-0.04}$ & 0.012  $^{+0.006 } _{-0.001 }$ & 9.9   $^{+  5.47 } _{-  0.33 }$ \\ \vspace{0.24cm}
HD 89263           & 1.52 $^{+0.23} _{-0.07}$ & 0.011  $^{+0.013 } _{-0.006 }$ & 561.8 $^{+194.6} _{-555.8}$ \\ \vspace{0.24cm}
%\\
\enddata
\end{deluxetable}

\section{Conclusion} \label{concl}
We have deployed a least-squares minimization technique to obtain the structure parameters of 60 $\delta$ Sct stars that show regular p-mode pulsation patterns. This method allows us to carry out spectrum fitting without prior mode identification. It is semi-automated in the sense that we fit very few modes in the trial-and-error effort, and mode identification comes as a byproduct of the fitting routine. We provided the first inferences of $M$, $Z$ ([Fe/H]) and age ($\tau$) for these stars.

We found the masses of most of these stars to be distributed around $1.6 M_\odot$, with the exception of two stars of lower mass (HD 2280, HD 10961: $\sim 1.4 M_\odot$) and one star with the highest mass (HD 290750: $2.1 M_\odot$). Metallicity of a significant fraction of the stars are found to hover around $Z = 0.010$ (or [Fe/H] = -0.23). In this sample, we also found a few stars having very low as well as very high values of metallicity. TIC 349645354 and KIC 8415752 have the highest metallicity, at $Z=0.024$. Finally, more than half of the stars turned out to be younger than 30 Myr, with the rest several hundreds of Myrs old. %Considering the fact that age inference for $\delta$ Sct stars is significantly limited by degeneracy, our results should be relied on for primarily qualitative inferences.

We had originally developed three machine learning based neural networks to carry out similar parameter inferences. Although we were unable to obtain reliable results, they still carry the potential of simpler as well faster parameteric fitting. On the synthetic data, our neural networks were able to infer $M$ and $Z$ much more accurately using \{$L,\Tf,\Dn,\eps$\} as inputs. However, age inferences were in general not as accurate as $M$ or $Z$. This is because, over the course of their MS evolution, stars cross their pre-MS trajectory, which make {\it all} their structure parameters degenerate in these two stages of evolution. A similar behaviour was demonstrated in \citealt{seismic_age}. Therefore, age inference is not expected to be confident.

With the exception of the parameter inference, neural networks allowed us to carry out an additional analysis -  we were able to determine the relative importance of different input quantities . We determined that $\Tf$ plays a critical role in the inference of all parameters. $\eps$ was found to have significance as important as $\log Z$ and other parameters like $M$, $\Dn$, and $L$, while constraining the age.

We observed that a longer pattern of radial modes (starting from $n = 1$) contains critical information about $\delta$ Sct structure. This set of frequencies may be treated as an essential substitute to \{$L,\Tf$\} as we found that they can constrain stellar parameters even without $L$ and $\Tf$. Doing the feature importance experiment, we found that $\nu_6$, $\nu_2$ and $\nu_3$ are the most significant radial modes.

Finally, we noticed that inclusion of dipole modes led to more precise determine $\delta$ Sct age. We used $L$, $\Tf$ and frequencies of both radial ($\ell = 0$) and dipole ($\ell = 1$) modes with radial order $n \in [4 - 7]$ as inputs to infer different parameters. The degeneracy problem in age was reduced while using these inputs. We therefore conclude that dipole modes act as independent quantities with which to constrain stellar parameters and they add supplementary information to radial modes and \{$\Dn,\eps$\}.

Our models lack the implementation of rotation and gravity darkening - which have significant impacts over stellar evolution and pulsation. While 5 km s$^{-1}$ fluctuation in rotation velocity can perturb the pulsation frequencies by $2-3\%$ (\citealt{Deupree_2011}), gravity darkening can cause as high as $\sim 50\%$ and $2.5\%$ departures in $L$ and $\Tf$ (\citealt{MESA2019}) respectively. Depending on these facts, our parameter inferences can be subject to modification once we introduce such phenomena. However, this is beyond our current scope and we look forward to study their impacts in a future project. Such a detailed analysis will help us in putting more realistic constrains over the ages of multiple $\delta$\,Sct stars observed in missions like \textit{TESS} and \textit{Kepler}.

%\begin{acknowledgments}
\section*{acknowledgments}
S.D. acknowledges SERB, DST, Government of India, CII and Intel Technology India Pvt. Ltd. for the Prime minister's fellowship and facilitating research. We  have performed all computations in the Intel Lab Academic Compute Environment using Intel\textsuperscript{\textregistered} Xeon\textsuperscript{\textregistered} Platinum 8280 CPU. MESA (\citealt{MESA2011}) and GYRE (\citealt{GYRE2013}) codes have been very useful in simulating the data, without which this project would not have been possible. We have used SciPy (\citealt{SciPy}), Matplotlib (\citealt{Matplotlib}), NumPy (\citealt{NumPy}), Keras (\citealt{keras}), and Tensorflow (\citealt{tensorflow}) like Python packages. This research made use of Lightkurve (\citealt{Lightkurve}), a Python package for Kepler and TESS data analysis. GNU parallel (\citealt{gnuprl}) has been used to do parallel computation. SJM was supported by the Australian Research Council (ARC) through Future Fellowship FT210100485. TRB was also supported by the ARC, through DP210103119 and FL220100117, and by the Danish National Research Foundation (Grant DNRF106) through its funding for the Stellar Astrophysics Centre (SAC). We thank Antonio Garc\'ia Hern\'andez for his suggestions regarding the impacts of rotation and gravity darkening. Finally we are thankful to the referee for his very useful comments.

\appendix
\restartappendixnumbering
\section{Inference for each star} \label{inf-real}
%%%%%%%%%%%%%%%%%%%%%%%%%%%%%%%%%%%%%%%%%%%%%%%%%%%%%%%%%%%%%
We show below in Figures \ref{inf-real-M} - \ref{inf-real-t}, $M$, $Z$, and $\tau$ inferences from method-4 for 60 $\delta$\,Sct stars taken from \citealt{nature}. The x-axis contains the IDs of the stars and the y-axis displays the inferred quantities.

\begin{figure*}[tb]%{\columnwidth}
    \centering
    \includegraphics[width=\textwidth]{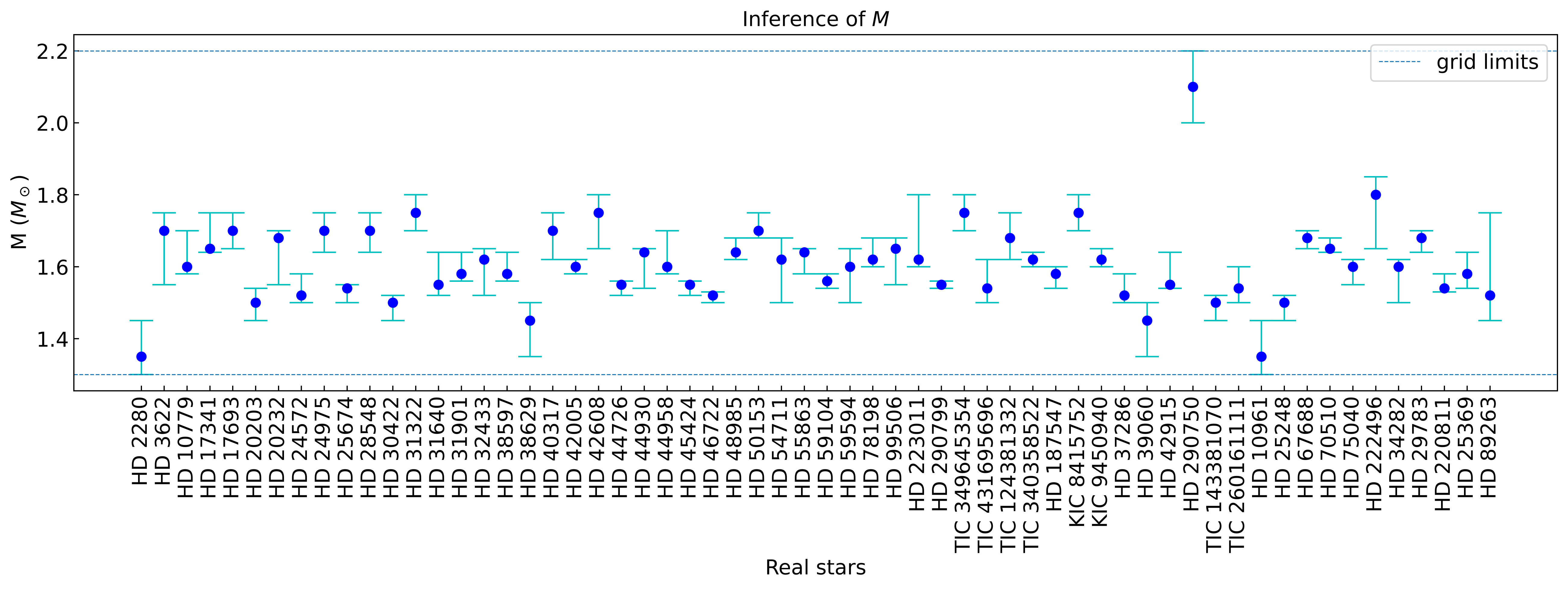}
    %\captionsetup{labelfont={color=blue, bf}}
    \caption{$M$ inference for 60 $\delta$ Scuti stars.}
    \label{inf-real-M}
\end{figure*}

\begin{figure*}[tb]%{\textwidth}
    \centering
    \includegraphics[width=\textwidth]{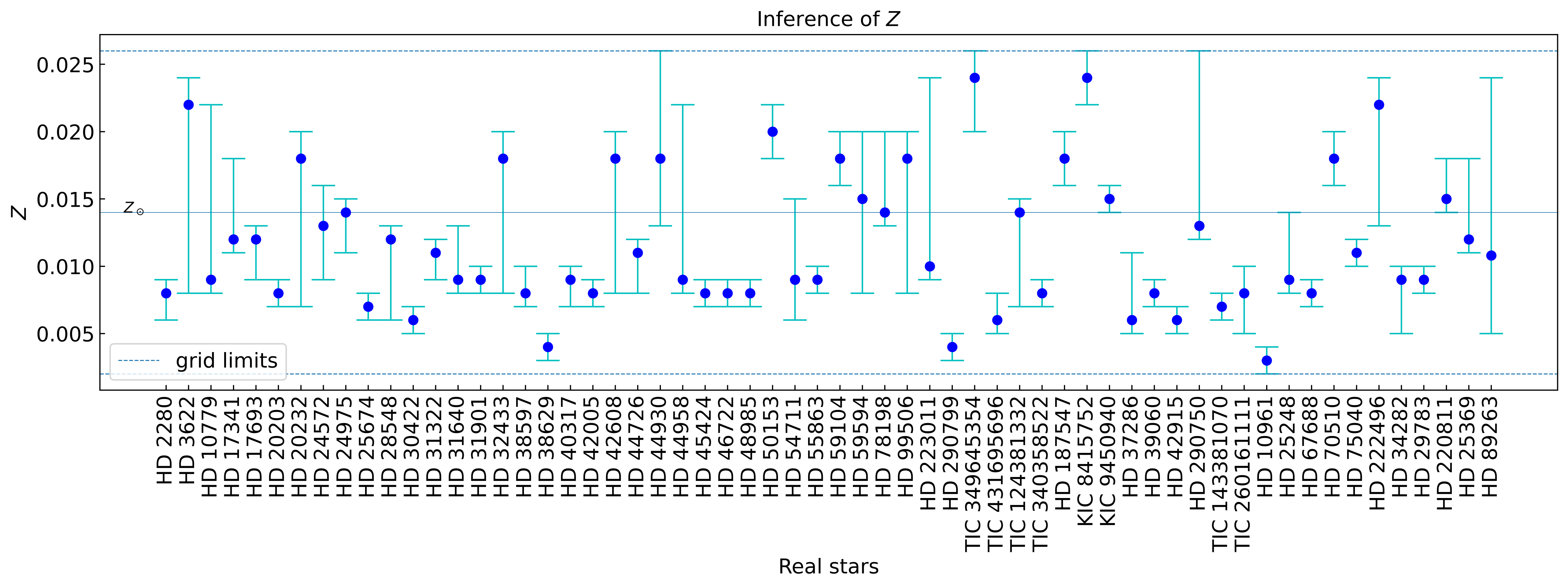}
    %\captionsetup{labelfont={color=blue, bf}}
    \caption{$Z$ inference for 60 $\delta$ Scuti stars.}
    \label{inf-real-Z}
\end{figure*}

\begin{figure*}[tb]%{\textwidth}
    \centering
    \includegraphics[width=\textwidth]{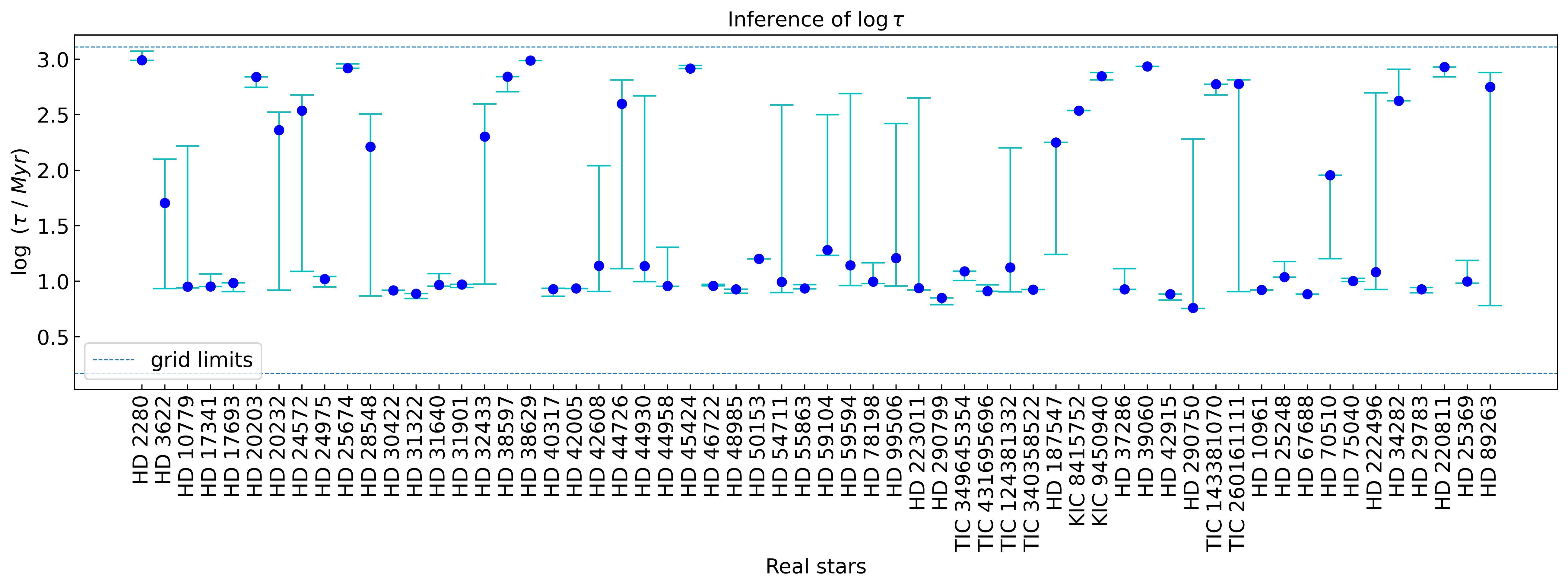}
    %\captionsetup{labelfont={color=blue, bf}}
    \caption{$\log_{10} (\tau / {\rm Myr})$ inference for 60 $\delta$ Scuti stars.}
    \label{inf-real-t}
\end{figure*}

\section{Group-wise inference accuracy}
%%%%%%%%%%%%%%%%%%%%%%%%%%%%%%%%%%%%%%%%%%%%%%%%%%%%%%%%%%%%%%%
For most of the stars, measured values of $L$ and $\Tf$ are available, although these quantities are not sufficient to constrain different stellar parameters. Additional independent quantities are expected to assist in constraining these structure parameters.

Asteroseismic quantities $\Delta\nu$ and $\eps$ are suitable for this purpose. But neither ($L, \Tf$) nor ($\Dn, \eps$) alone can accurately infer the stellar parameters. However, when combined, these parameters can efficiently constrain those stellar parameters.

\begin{figure*}[tb]
    \includegraphics[width=\textwidth]{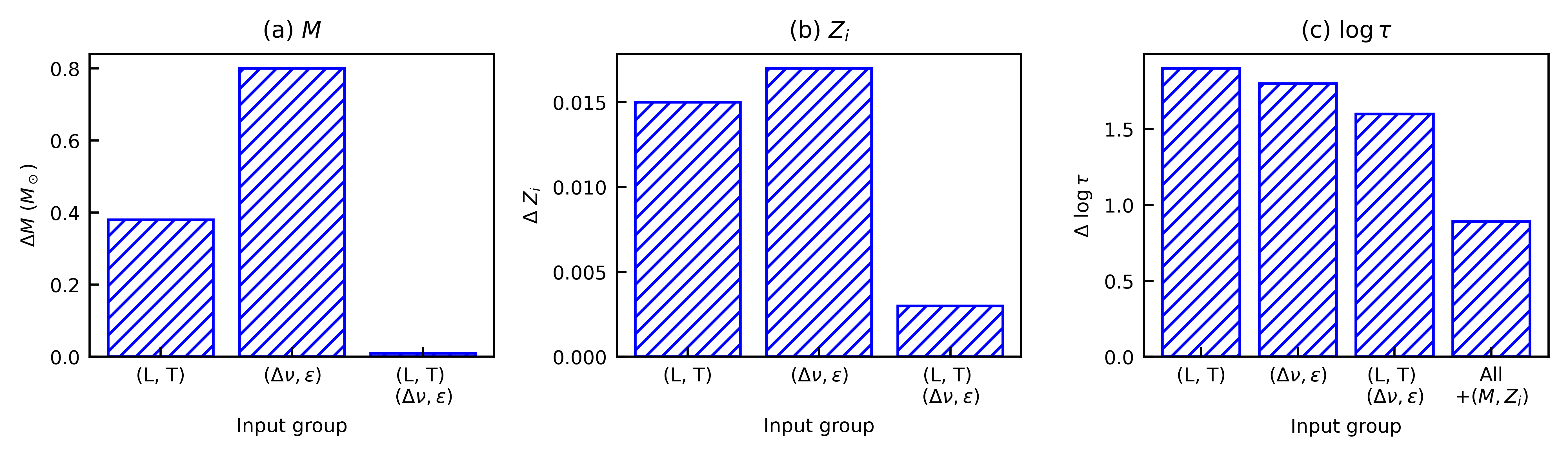}
    %\captionsetup{labelfont={color=blue, bf}}
    \caption{Maximum absolute errors between network prediction and true values, for a different group of input quantities shown in the x-axis. Figures for prediction of (a) $M$ in $M_\odot$, (b) $Z$ and (c) $\log\tau$, where $\tau$ is in Myr. \label{fig:group-error}}
\end{figure*}

We considered \{$L, \Tf, R$\} as inputs, trained the networks with these, and measured the maximum absolute errors between true values and network predictions (over the validation samples). Similarly, we took asteroseismic quantities \{$\Dn, \eps$\} and repeated this validation process. Finally, we combined all of them and again measured the maximum absolute error associated with the predictions. During age inference, we supplemented $M$, $Z$ like quantities and also assembled the values of validation errors. We visualize all of these errors in Figure \ref{fig:group-error}, which emphasizes that validation error decreases drastically when observables and asteroseismic quantities are simultaneously considered as inputs.

In Table \ref{t:group-R}, we show the Pearson-R correlation coefficients (\citealt{Pearson-R}) between network predictions (over validation samples) and corresponding true values. It is a measure of inference accuracy, i.e., correlation between true and inferred values. The Pearson-R coefficient can have values between -1 and +1. The higher the R coefficient, the higher the correlation between true values and network prediction. This Pearson-R correlation coefficient between the two sets of measurements \{p\} and \{q\} is calculated by equation \ref{eq:Pearson-R}.
\begin{equation} \label{eq:Pearson-R}
    R = \dfrac{ \sum_i(p_i-\overline{p})(q_i-\overline{q})}{ \sqrt{ \sum_i(p_i-\overline{p})^2 \sum_i(q_i-\overline{q})^2 } }
\end{equation}
where, $\overline{p}, \overline{q}$ are the means of the measurements.

Adding ($M, Z$) as additional inputs increases the Pearson-R coefficient by only 0.01. However, from Figure \ref{fig:group-error} (c), it is evident that ($M, Z$) inputs actually assist in reducing the absolute error associated with $\log \tau$ prediction.

\begin{table*}[htb]
    \centering
    \begin{tabular}{ccccc}
    \hline \hline
                    & \{$L,\Tf$\} & \{$\Dn,\eps$\} & \{$L, \Tf$,   & \{$L, \Tf$, \\
                    &             &                & $\Dn, \eps$\} & $\Dn, \eps$, \\
                    &             &                &               & $M, Z$\} \\ \hline
        $M$         & 0.93        & 0.86           & 0.99          & \\
        $\log Z$  & 0.77        & 0.85           & 0.99          & \\
        $\log \tau$ & 0.67        & 0.78           & 0.98          & 0.99 \\ \hline
    \end{tabular}
    \caption{Pearson-R correlation coefficients between network predictions and true values, for different groups of input quantities.}
    \label{t:group-R}
\end{table*}

\section{supplementary resources}
%%%%%%%%%%%%%%%%%%%%%%%%%%%%%%%%%%%%%%%%%%%%%%%%%
This article provides a supplementary file demonstrating the stellar model fittings to observations of all stars in the sample.

\bibliographystyle{aasjournal}
%\bibliography{main.bib}
%\input{main.bbl}

\end{document}